\documentclass{article}[12pt,a4paper]

\usepackage{geometry}
\date{}

\usepackage{setspace}
\usepackage{hyphenat} 
\usepackage[latin1]{inputenc} 
\usepackage{graphicx}
\usepackage[usenames]{color}
\usepackage{multirow}
\usepackage[FIGBOTCAP]{subfigure}
\usepackage{hyphenat} 
\usepackage[latin1]{inputenc} 
\usepackage{graphicx}
\usepackage[usenames]{color}
\usepackage{amssymb,latexsym,amsmath,srcltx,amsfonts,color}
\usepackage[latin1]{inputenc}
\usepackage{accents}
\usepackage{multirow}
\usepackage[FIGBOTCAP]{subfigure}
\usepackage{empheq,amsmath} 
\usepackage{subeqnarray}

\RequirePackage[OT1]{fontenc}

\DeclareGraphicsExtensions{.pdf,.png,.jpg,.tif,.ps,.eps}

\usepackage[pdftex,bookmarks=true]{hyperref}

\newcommand{\bfsigma}{\mbox{\boldmath $\sigma$}}
\newcommand{\bftheta}{\mbox{\boldmath $\theta$}}

\newcommand{\bfmu}{\mbox{\boldmath $\mu$}}

\newcommand{\bfPhi}{\mbox{\boldmath $\Phi$}}

\begin{document}

\titlepage
\setcounter{page}{1}
\title{\bf Mixture models applied to heterogeneous populations}
\author{{Carolina Valani Cavalcante\footnote{Escola Nacional de Ci\^encias Estat\'{\i}sticas (ENCE), Rio de Janeiro, RJ, Brazil. 
Email: carolinavalanic@gmail.com} and Kelly Cristina M. Gon\c{c}alves\footnote{Departamento de Estat\'{\i}stica, Universidade Federal do Rio de Janeiro
(UFRJ), RJ, Brazil. Email: kelly@dme.ufrj.br}} }
\maketitle \vspace*{-30pt}
\begin{center}
\end{center}

\begin{abstract}
    Mixture models provide a flexible representation of heterogeneity in a finite number of latent classes. From the Bayesian point of view, 
    Markov Chain Monte Carlo methods provide a way to draw inferences from these models. In particular, when the number of subpopulations
    is considered unknown, more sophisticated methods are required to perform Bayesian analysis. The Reversible Jump Markov Chain Monte 
    Carlo is an alternative method for computing the posterior distribution by simulation in this case. Some problems associated with the 
    Bayesian analysis of these class of models are frequent, such as the so-called ``label-switching" problem. However, as the level of 
    heterogeneity in the population increases, these problems are expected to become less frequent and the model's performance to improve. 
    Thus, the aim of this work is to evaluate the normal mixture model fit using simulated data under different settings of heterogeneity and 
    prior information about the mixture proportions. A simulation study is also presented to evaluate the model's performance considering 
    the number of components known and estimating it. Finally, the model is applied to a censored real dataset containing antibody levels of 
    Cytomegalovirus in individuals.
\end{abstract}




\section{Introduction}\label{intro}

Mixture models are noted for their flexibility in modeling complex data and are widely used in the statistical literature (see 
\cite{mclachlan2004finite}). These models provide a natural framework for modeling heterogeneity in a population. Moreover, due to the 
large class of functions that can be approximated by mixture models, they are attractive for describing non-standard distributions and have 
been adopted in many areas, such as genetics, ecology, computer science, economics, biostatistics and many others. For instance, as stated in 
\cite{jordan2004graphical}, in genetics, location of quantitative traits on a chromosome and interpretation of microarrays are both related to 
mixtures, while, in computer science, spam filters and web context analysis start from a mixture assumption to distinguish spams from regular 
emails and to group pages by topic, respectively. 

Statistical analysis of mixtures is not straightforward and the Bayesian paradigm has been particularly suited for this purpose. 
This framework allows the complicated structure of a mixture model to be decomposed into a set of simpler structures through the use of hidden 
or latent variables. According to \cite{richardson1997bayesian}, when the number of components is unknown, the Bayesian paradigm is the only 
sensible approach to its estimation. Thus, the Bayesian approach has enabled mixture models to become increasingly popular in many areas. In
real applications, the number of components can have important implications regarding the problem, so it has to be well specified or estimated, 
although practitioners usually have little theoretical guidance. On the other hand, even if prior theory suggests a particular number of 
components, it may not be possible to reliably distinguish between some of the components. In some cases additional components may simply reflect the presence of 
outliers in the data.

When the number of subpopulations is assumed to be known, Markov Chain Monte Carlo methods (MCMC) can be used for Bayesian estimation of the 
subpopulation parameters. Nevertheless, this method, as originally formulated, requires the posterior distribution to have a density with 
respect to some fixed measure. When the number of components is considered unknown, i.e., the size of the parameter space is also a parameter, 
a problem with variable dimension appears, so MCMC cannot be used alone and more sophisticated methods are required to perform 
the Bayesian analysis. One alternative in this case is the approach based on Reversible Jump MCMC (RJMCMC), which was first proposed by 
\cite{green1995reversible} and applied in univariate normal mixture models with unknown numbers of components by \cite{richardson1997bayesian}. 
The method basically consists of jumps between the parameter subspaces corresponding to different numbers of components in the mixture.

While MCMC provides a convenient way to draw inference from complicated statistical models, there are still many, perhaps under appreciated, 
problems associated with the MCMC analysis of mixtures. These problems are mainly caused by the nonidentifiability of the components under 
symmetric priors, which leads to the so called label-switching problem in the MCMC output, discussed in \cite{jasra2005markov}. The term 
describes the invariance of the likelihood function under relabelling of the mixture components, which can cause the posterior distribution of 
the parameters to be highly symmetric and multimodal. Therefore, the component labels are mixed up
and cannot be distinguished from each other. As a result, the usual 
practice of summarizing joint posterior distributions by marginal distributions, and for instance estimating parameters by their posterior 
mean is often inappropriate, because the marginal on the parameters for all components is identical and the posterior expectation for the
parameters is identical too. A frequent response to this problem is to remove the symmetry by using artificial identifiability constraints. 
This and other alternative approaches to this problem are described by \cite{stephens2000dealing}. 


 The aim of this work is to review and discuss some aspects in the inference of mixture models, in particular the normal mixture 
models, when applied to heterogeneous populations under the Bayesian approach. The main purpose is to evaluate the model's performance in 
different settings of heterogeneity, prior information for mixture weights and when considering the number of components known or unknown.

Most previous works, as \cite{jasra2005markov}, in general, evaluate a possible improvement in the label-switching problem choosing a more 
appropriate relabelling algorithm. However, the aim here is to verify if the label-switching phenomena generally persists when the 
subpopulations are not well separated and how the level of heterogeneity of the population affects the results. Furthermore, we evaluate the 
label-switching phenomenon by assuming more informative prior distributions for the mixture weights and show that in some cases it can be 
more efficient than use different relabelling algorithms known in the literature.

Although works as \cite{nobile2004posterior} and \cite{richardson1997bayesian} carried out some studies about the influence of hyperparameters 
specification on posterior distribution of other parameters, different from this paper, they did not connect the prior distribution of the weights with the 
label-switching problem, which is our main result.

On the other hand, previous works that compare the results obtained under RJMCMC with MCMC, as \cite{richardson1997bayesian} for the univariate 
case and \cite{dellaportas2006multivariate} for the multivariate one, use as a diagnostic tool mainly the predictive density and the posterior 
estimates of $k$. In this work we also evaluate the model's performance under a simulation study in order to verify frequentist properties of 
the Bayes' estimators produced from each method.

The paper is organized as follows. Section \ref{model} presents the general definition of a mixture model and discusses some aspects of 
inferences. A simulation study for assessing the estimation of model parameters under different levels of heterogeneity is presented in 
Section \ref{simul_study}. Additionally, a prior sensitivity analysis of the mixture proportions is presented in order to see its effect under
label-switching phenomena. We also discuss the model 
fit when the number of components is known and unknown. In Section \ref{real_data} the performance of the 
method is assessed through an application to a left-censored real dataset. Thus, we discuss briefly the inference in this case. Finally, Section \ref{sec:concl} presents some conclusions and suggestions 
for further research.

We particularly used the \cite{teamR} package called $\mathtt{mixAK}$, proposed by \cite{komarek2009new}, with routines to compute the posterior distribution through MCMC and RJMCMC for multivariate right, left and interval-censored observations. Appendix A presents the main code that can be used to perform the experiments presented here.

\section{Finite mixture models}\label{model}

The basic mixture model for independent scalar or vector observations $Y_i$, $i=1,\dots,n,$ is a convex combination given by:
 \begin{eqnarray}\label{mixt}
 Y_i\sim \sum_{j=1}^k{w_jf(\cdot\mid \bftheta_j)},\, i=1,\dots,n,
 \end{eqnarray}
where $f(\cdot\mid \bftheta)$ is a given parameter family of densities indexed by a scalar or a vector $\bftheta$. In general, the objective 
of the analysis is to make inferences about the unknowns: the number of components, $k$; the parameters $\bftheta=(\bftheta_1,\dots,\bftheta_k)$
with $\bftheta_j$ being specific to component $j$; and the components' weights, ${\bf w}=(w_1,\dots,w_k)$, $0<w_j<1$, $\sum_{j=1}^k{w_j}=1$. 
Let ${\bf \Phi} = ({\bf w},\bftheta,k)$ be the parametric vector of the model (\ref{mixt}).

For an observed random sample ${\bf y} = (y_{1}, y_{2}, \dots, y_{n})'$, the likelihood function of ${\bf \Phi}$ is given by:
\begin{eqnarray}\label{likel1}
p({\bf y}|\bfPhi) = \prod_{i=1}^{n} \sum_{j=1}^{k} w_{j} f(y_{i}|\bftheta_{j}). 
\end{eqnarray}
The likelihood function leads to $k^n$ terms, which creates computational difficulty.

A context in which the model (\ref{mixt}) can arise and we are interested in this paper is when we postulate a heterogeneous population 
consisting of heterogeneous groups $j=1,2,\dots,k$ of sizes proportional to $w_j$, from which a random sample is drawn. The label of the group 
from which each observation is drawn is unknown, so it is natural to regard the group label $z_i$, for 
the $i$-th observation as a latent variable and rewrite (\ref{mixt}) as the following hierarchical model: for $i=1,\dots,n$, $j=1,\dots,k,$
\begin{eqnarray}\label{mixt2}
Y_i\mid \bftheta_j,z_i=j\sim f(\cdot\mid \bftheta_{j}), \mbox{ with } P(z_i=j)=w_j.
\end{eqnarray}

By integrating ${\bf z}=(z_1,\dots,z_k)$ out from (\ref{mixt2}) we return to model (\ref{mixt}). The formulation given by (\ref{mixt2}) is 
convenient for interpretation and calculation. Latent indicator variables usually leads to an efficient simulation algorithm that 
quickly focuses on the modes of the posterior distribution, thus it will contribute to reducing the computational effort.

A Bayesian approach to inference requires the specification of a prior distribution $p(.)$ for the parameters of the mixture model (\ref{mixt}).
In particular, prior elicitation is an important step. According to \cite{roeder1997practical}, in a mixture content, assuming a
non-informative prior yield improper posterior distributions. Since there is always the possibility that no 
observations are allocated to one or more components, standard choices of independent improper non-informative prior distributions for the 
component parameters cannot be used. \cite{richardson1997bayesian} proposes an alternative in this case based on keeping the simple independence 
and define weakly informative priors, which may or not be data dependent.

The mixture model in (\ref{mixt2}) is invariant to permutation of the labels $j=1,\dots,k$. Some implications of this for likelihood analysis 
are discussed by \cite{redner1984mixture}. If  we  have  no  prior information  that  distinguishes  between  the  components  of  the  mixture,
so  the  prior distribution $p(.)$ is the same for all permutations of $\bftheta$, then the posterior distribution will also be symmetric and, there will be $k!$ 
symmetric modes in the posterior distribution. Then, the marginal posterior distributions for the parameters will 
be identical for each mixture component and the posterior means of all the components are the same, thus they are poor estimates of these 
parameters. Thus, if the problem is not handled properly, the ergodic average of the MCMC samples is not appropriate for the estimation of the 
parameters. 


There are some suggested solutions to this problem, see \cite{stephens2000dealing} for details. One common response to the label-switching 
problem is to impose an identifiability constraint on the parameter space. This breaks the symmetry of the prior and thus, of the posterior 
distribution of the parametric vector. For example, we can impose an ordering constraint on $\theta_j$'s, such as 
$\theta_1<\theta_2<\dots<\theta_k$, if it is a scalar. 

\subsection{Inference}\label{sec:inf}

Since we are in a Bayesian framework, the inference consists of obtaining the posterior distribution of the parametric vector $\bfPhi$ of
model (\ref{mixt2}). In general, this joint distribution cannot be obtained in closed form. One alternative, which is often used and is 
feasible to implement, is to generate samples from the marginal distributions of the parameters based on the MCMC algorithm. A comprehensive
Bayesian treatment using MCMC methods was presented in \cite{diebolt1994estimation} for finite mixture models.

Nevertheless, this method, as originally formulated, requires the posterior distribution to have a 
density with respect to some fixed measure. Thus, in the mixture context, the method can only be applied when the number of components $k$ in 
the model (\ref{mixt2}) is considered known. 

However, the number $k$ is rarely known, and setting an incorrect value can bring important consequences to the posterior distribution. 
For instance, on an extremal case, a mixture model with only one component will roughly impose zero density to some ranges of the 
density, while a mixture with the number of components equal to the number of observations, simply will over-fit the data, creating unnecessary 
clusters. Thus, a halfway solution offers a trade off between these two solutions, providing a good fit of the data, modeling well all ranges of the density,
without a wrong large number of parameters to estimate. Other times, the target of the study is exactly the estimation of $k$. The approach based on RJMCMC is an alternative in this case, as proposed in 
this context by \cite{richardson1997bayesian}. It operates on the augmented parameter space, where the allocation variables ${\bf z}$ are included as unknown parameters. 

The method basically consists of jumps 
between the parameter subspaces corresponding to different numbers of components in the mixture, after updating them. If the current model is a 
mixture with $k>1$ 
components, then it is usual to reduce the searching strategy to moves that either preserve the number of components, or lead to a mixture with
$k-1$ or $k+1$ components. The idea is then to supplement each of the spaces with adequate artificial spaces
in order to create a bijection between them, most often by augmenting the space of the smaller model. Jumps are achieved by adding new 
components, deleting existing components, and splitting or merging these. These moves are randomly chosen and after being drawn, it is 
necessary to make corresponding changes to $(\bftheta,{\bf w})$. 

One can assess convergence for each mixture component parameters, however, as mentioned before, label-switching issues
lead to a type of poor mixing in the mixture component-specific parameters, which may not impact convergence and mixing of the induced
predicitive density. In particular, due to exchangeability of the mixture components, its marginal posterior distributions is identical
for all subgroups and hence the chains for each of the mixture component-specific parameters have the same target distribution. For instance,
in a normal mixture model with $k=2$ components, suppose one mixture component located at $\mu_1=0$ and the other at $\mu_2=-1$. The Gibbs sampler
for $\mu_1$ should then randomly jump between values close to 0 and values close to -1 if mixing is good. The posterior for $\mu_i$, $i=1,2$
has a multimodal form with one mode close to 0 and one mode close to -1. If these modes are well separated and there is a region of low 
probability density between the modes, then the Gibbs sampler will remain stuck for long intervals in one mode. A well-mixing Gibbs sampler
should switch between these modes often, but in practice for well separated components, it is common to remain stuck in one labeling across
all the samples are collected.

For those reasons, it is not appropriate to simply calculate posterior summaries based on the posterior draws of the parameters.
In the two-component illustration, one would obtain the same posterior mean for $\mu_1$ and $\mu_2$ if the Gibbs sampler mixed well
enough and sufficiently many samples were drawn. If we could relabel the samples so that after relabelling all the samples of 
$(\mu_1,\sigma_1^2)$ correspond to the component at $\mu=0$ and all the samples of $(\mu_2,\sigma_2^2)$ correspond to the component
at $\mu=-1$, then we could calculate posterior summaries of the mixture component-specific parameters in the standard manner. This relabelling 
can be done using postprocessing algorithms or with constraints in the prior in an attempt to make the separate mixture components distinguishable.
In this paper, we will restrict imposing an increasing order to the mixture means in the prior distribution, so that the higher indexed
components have higher means.

However, for any given dataset, many identifiability constraint choices can
be ineffective in removing the symmetry from the posterior distribution. For instance, in multivariate cases it is not at all clear in most
cases what type of restriction is appropriate. Even in the univariate case, it may be that we need multiple components with similar means
but different variances to provide a good fit to the data. If the means are close together, then label-switching can occur even if we place
a strict order restriction on the means and hence the restriction does not fully solve the label ambiguity problem.



\subsection{Normal mixture model}

In this work we are particularly interested in the univariate normal case presented in \cite{richardson1997bayesian}, so $\bftheta_j$ in
(\ref{mixt}) becomes a vector with expectation and variance parameters $(\mu_j,\sigma_j^2)$. The model is stated below: for $i=1,\dots, n$ and 
$j=1, \dots, k$,
\begin{eqnarray}\label{modelo}
\begin{array}{rl}
Y_{i}\mid \mu_j,\sigma^2_j, z_{i} = j &\sim \mbox{ Normal }(\mu_{j}, \sigma^{2}_{j}),\\
P(z_{i} = j) &= w_{j}.\\
\end{array}
\end{eqnarray}

Assuming that the parameters in $\bfPhi$ are prior independent and identically distributed and that $k$ is unknown, the prior distribution is 
given by: for  $i=1,\dots, n$ and $j=1, \dots, k$,
\begin{eqnarray}\label{prior}
\begin{array}{rl}
{\bf w} &\sim \mbox{ Dirichlet }(\gamma), \\
\mu_{j} &\sim \mbox{ Normal }(\mu_{a}, \sigma^{2}_{a}), j=1, \dots, k, \\ 
\sigma^{-2}_{j} &\sim \mbox{ Gamma }(\alpha, \beta), j=1, \dots, k, \\ 
\beta &\sim \mbox{ Gamma }(g, h), \\
k &\sim \mbox{ Uniform }\{1, k_{max}\},
\end{array}
\end{eqnarray}
where $\mbox{Dirichlet }(a)$ generically denotes the symmetric Dirichlet distribution with parameter $a$. The symmetric Dirichlet distributions
are often used, since there typically is no prior knowledge favoring one component over another. Since all elements of the parameter vector 
have the same value, the distribution is alternatively parameterized by a single scalar value $a$. $\mbox{Gamma }(a,b)$ represents
the gamma distribution with mean $a/b$ and variance $a/b^2$ and $\mbox{Uniform }\{a,\dots,b\}$ is the uniform distribution defined on the integers
$\{a,\dots,b\}$. Moreover, for identifiability, we can use for example that the $\mu_j$ are in increasing numerical order, thus the joint prior 
distribution of $\bfPhi$ is $k!$ times the product of their marginal prior distributions.

In this paper, as done in \cite{richardson1997bayesian} we consider the Bayesian estimation in the set-up where we do not have strong prior
information about the mixture parameters. Thus, we use weakly informative priors and the default hyperparameter choices can be seen with 
further details in \cite{richardson1997bayesian}. 


As mentioned in Subsection \ref{sec:inf}, in this work the unique labeling is achieved by imposing a restriction on $\mu_j$. We use that in which the $\mu_j$ are in increasing 
numerical order; so the joint prior distribution of the parameters is $k!$ times the prior density, restricted to the set 
$\mu_1<\mu_2<\dots<\mu_k.$ When the means are well separated, labels of the realizations from the posterior by ordering 
their means generally coincide with the population ones. As the separation gets small, label-switching can occur. This problem can be also 
minimized by  choosing to order other parameters of the mixture components, for example, the variance, weights or some combination of all 
three parameters.

\section{Simulation study}\label{simul_study}

To assess the convergence of the MCMC and RJMCMC estimation, we generated some samples under different settings. We obtained samples from the 
posterior distributions of the model parameters, supposing $k$ known and estimating it. The population estimates were then compared with the 
true values to evaluate the model's performance. The aim was to evaluate the performance of the normal mixture model varying the level of 
heterogeneity and the prior information elicited for the mixture proportions. Furthermore, we also compared the results obtained under each simulation method considered, MCMC and RJMCMC.

\subsection{Assessment of RJMCMC and MCMC under different scenarios}\label{sens}
\begin{figure}[h!]
\begin{center}
\vspace{-0.7 cm}\begin{tabular}{cc}
\hspace{-0.5 cm}\subfigure[Heterogeneous]{\includegraphics[scale=0.5]{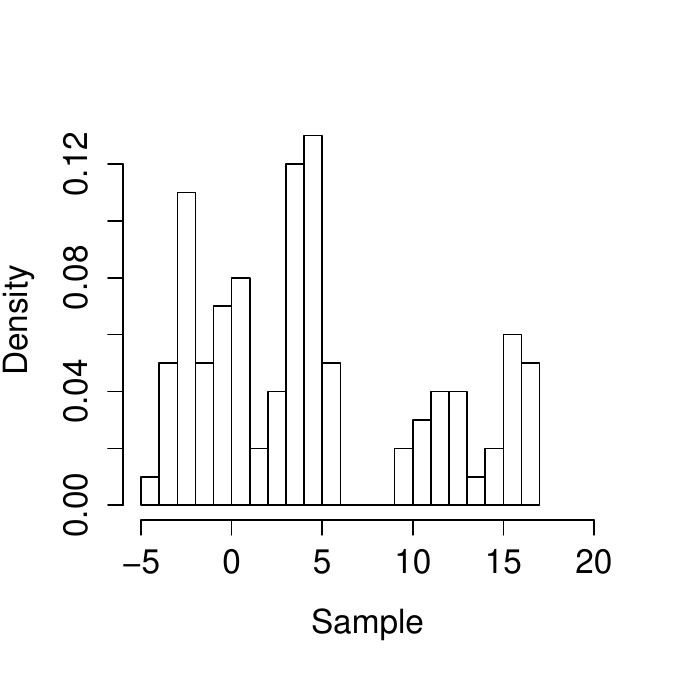}}\subfigure[Homogeneous]{\includegraphics[scale=0.5]{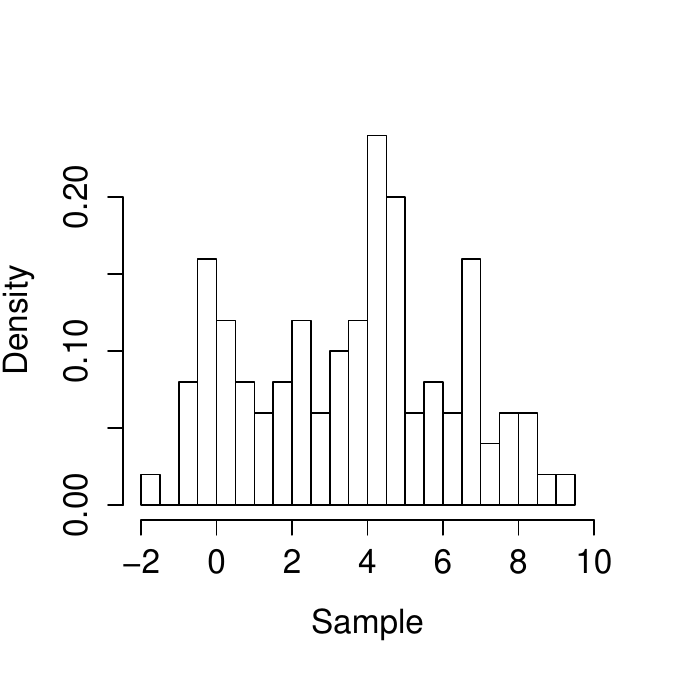}}
\end{tabular}
\end{center}
\vspace{-0.3 cm}\caption{{\it Histograms with the distribution of the samples generated.}}\label{data1}
\end{figure}

To check the convergence of the RJMCMC and MCMC estimations, we generated one sample with $n = 100$ observations under two levels of 
heterogeneity, the first one with well-separated groups, which we call the more heterogeneous sample, and the other with groups less well 
separated, which represents the more homogeneous one. In both scenarios we fixed $k=5$, $\bfsigma^2=(\sigma^2_1,\dots,\sigma^2_5)=(0.22,1.95,0.92,0.74,1.13)$ and 
${\bf w}=(w_1,\dots,w_5)=(0.17,0.21,0.34,0.12,0.16)$. With these values fixed for ${\bf w}$, we expected to have groups with a reasonable 
number of observations, so we did not consider scenarios with groups outliers. The heterogeneous scenario was obtained by fixing 
$\bfmu=(\mu_1,\dots,\mu_5)=(-3,0,4,11,16)$ and the homogeneous by setting $\bfmu=(\mu_1,\dots,\mu_5)=(0, 2, 4, 6, 8)$. Figure \ref{data1} 
presents the distribution of both datasets generated. The aim of this study is to verify how the level of heterogeneity of the population affects the results,
mainly regarding the label-switching problem.

The prior distribution considered are described in (\ref{prior}), and we elicited the prior for $\mu_j$ and $\sigma_j^2$ using the 
same idea of weakly informative prior suggested by \cite{richardson1997bayesian}. Moreover, we sort all components according to increasing 
order of magnitude of the posterior means to avoid the label switching phenomenon. 

First, we assumed $k$ unknown and in its prior distribution 
presented in (\ref{prior}) we assumed $k_{max}=10$, thus RJMCMC was used to obtain samples from the posterior distribution. 
We also did a brief prior sensitivity 
analysis assuming two values of the Dirichlet prior distribution for each dataset: 
$\gamma\in\{1,4\}$ for the heterogeneous case and $\gamma\in\{1,4,10\}$ for the homogeneous case. To assume $\gamma=1$ is equivalent to assuming 
a uniform distribution over all points in its support. On the other hand, the parameter value above $1$ gives some information that all 
sample proportions in subpopulations are similar to each other.

\begin{figure}[h!]
\begin{center}
\vspace{-0.7 cm}\begin{tabular}{cc}
\hspace{-0.7 cm}\subfigure[Heterogeneous]{\includegraphics[scale=0.4]{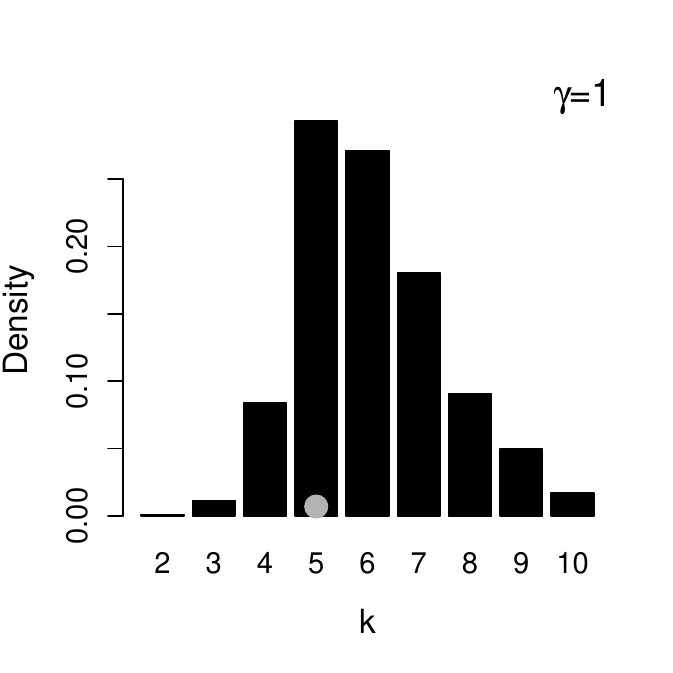}\includegraphics[scale=0.4]{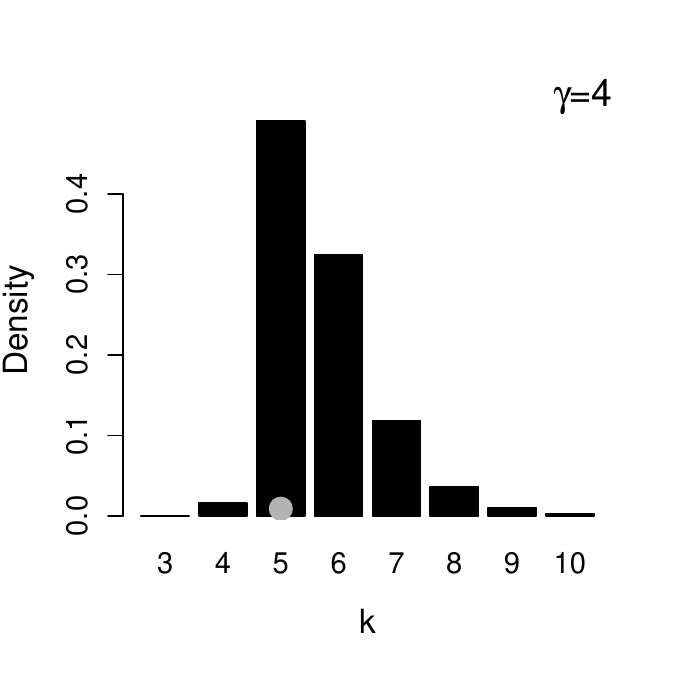}}\vspace{-0.4 cm}\\
\subfigure[Homogeneous]{\includegraphics[scale=0.4]{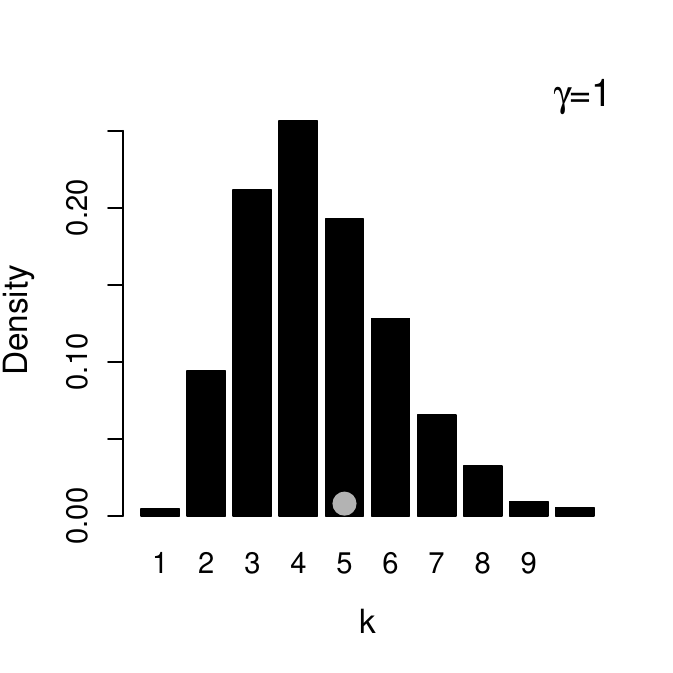}\includegraphics[scale=0.4]{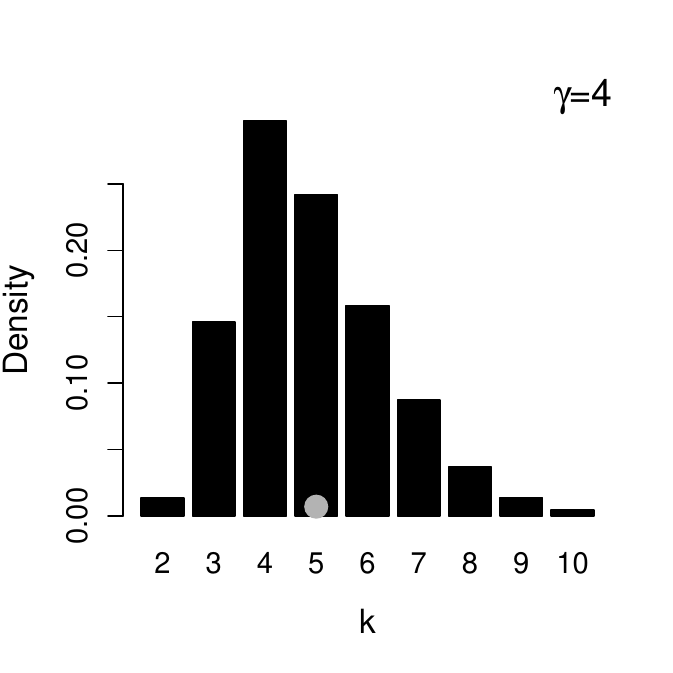}\includegraphics[scale=0.4]{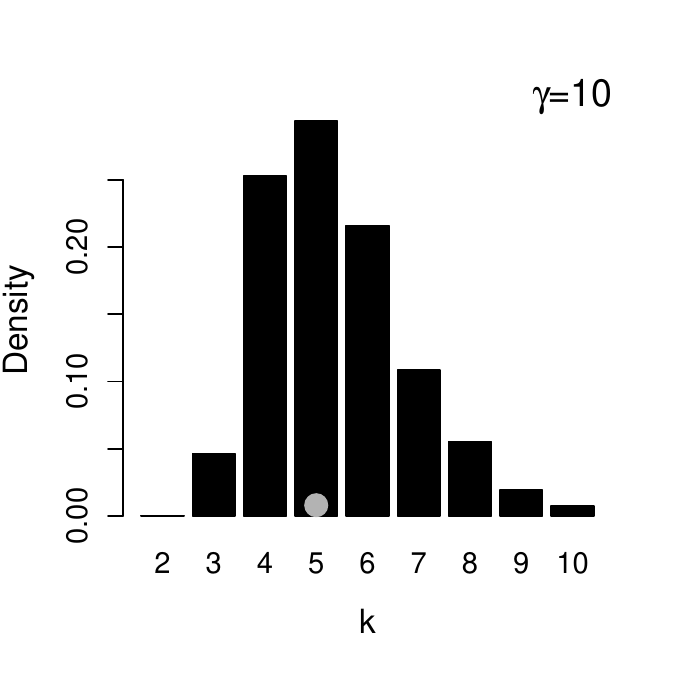}}
\end{tabular}
\end{center}
\vspace{-0.3 cm}\caption{{\it Posterior densities of the parameter $k$ for each value of $\gamma$ considered in the prior distribution
of the mixture proportions. The gray point represents the true value fixed in the
simulation.}}\label{kpost1}
\end{figure}

For the RJMCMC simulations, we generated, respectively for the homogeneous and heterogeneous cases, 350,000 and 70,000
samples from the posterior distribution, discarded the first 10,000 and 20,000, and then thinned the 
chain by taking every 10th sample value. Figure \ref{kpost1} displays the histogram with the posterior 
densities of $k$ for some values of $\gamma$.
It should be noted that for the heterogeneous case the parameter $k$ is well estimated, but when $\gamma=4$, the estimate is more accurate. 
On the other hand, $k$ is underestimated when assuming $\gamma=4$ with the homogeneous sample. The same happens with $\gamma=1$ or any value 
less than $4$. In this case, when $\gamma=10$ the value of $k$ is well estimated.


\begin{figure}[h!]
\begin{center}
\vspace{-0.7 cm}\begin{tabular}{l}
\hspace{-0.5 cm}\subfigure[Heterogeneous - RJMCMC]{\includegraphics[scale=0.33]{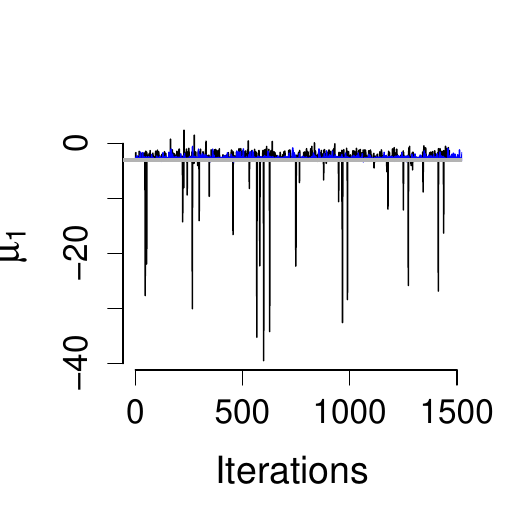}\hspace{-0.3 cm}\includegraphics[scale=0.33]{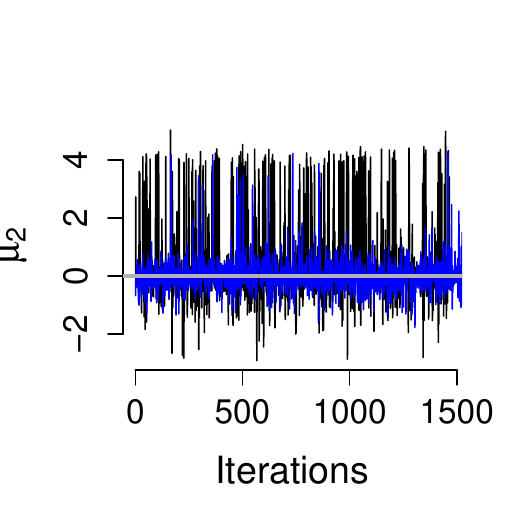}
\hspace{-0.3 cm}\includegraphics[scale=0.33]{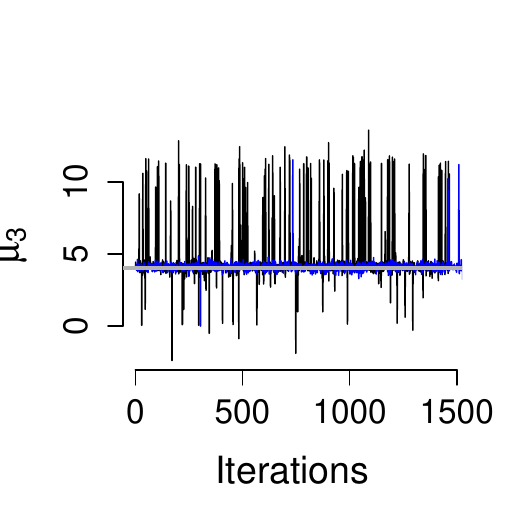}\hspace{-0.3 cm}\includegraphics[scale=0.33]{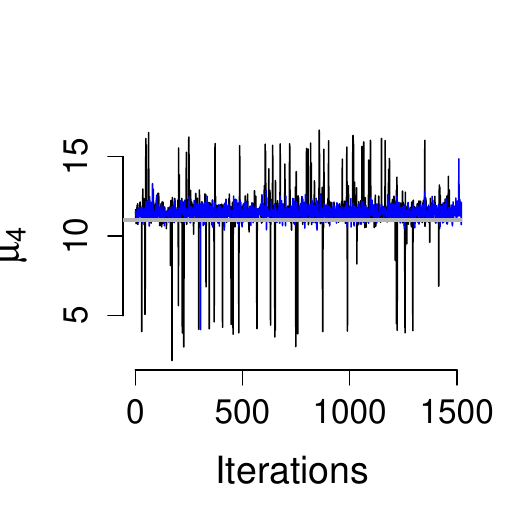}\hspace{-0.3 cm}\includegraphics[scale=0.33]{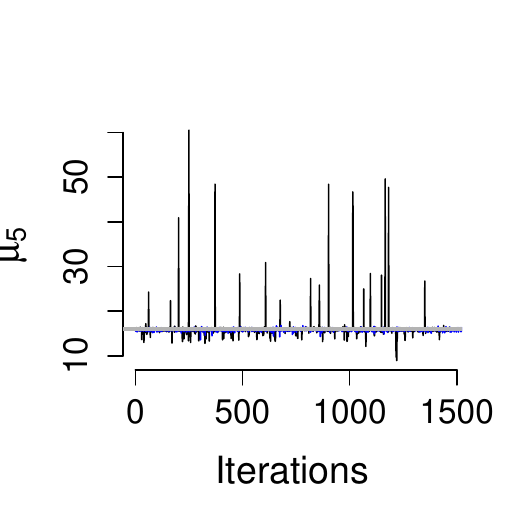}}\\
\hspace{-0.5 cm}\subfigure[Heterogeneous - MCMC]{\includegraphics[scale=0.33]{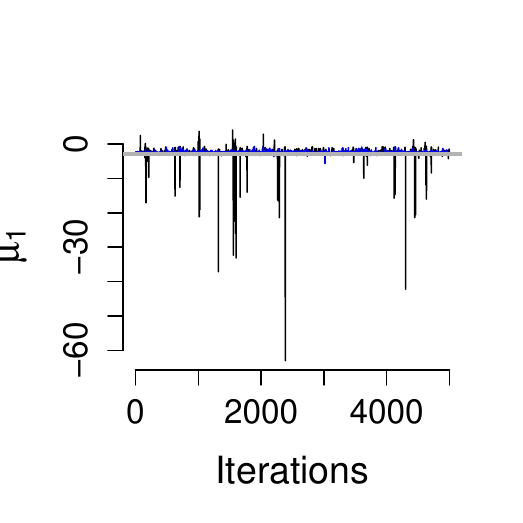}\hspace{-0.3 cm}\includegraphics[scale=0.33]{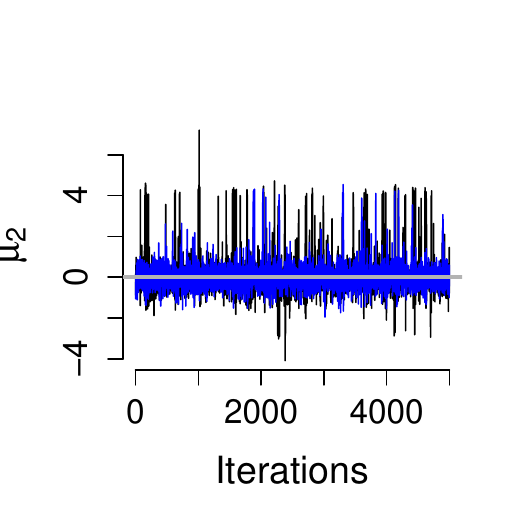}
\hspace{-0.3 cm}\includegraphics[scale=0.33]{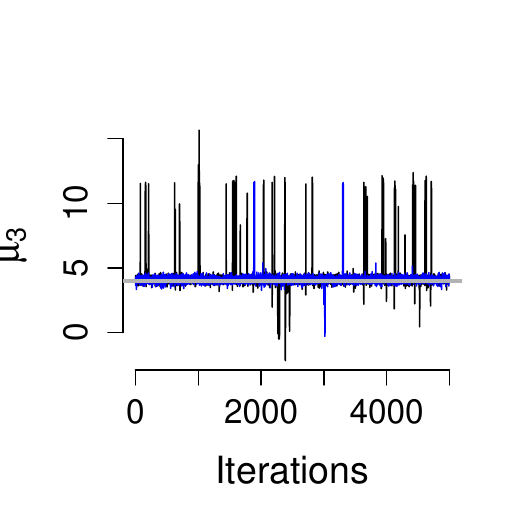}\hspace{-0.3 cm}\includegraphics[scale=0.33]{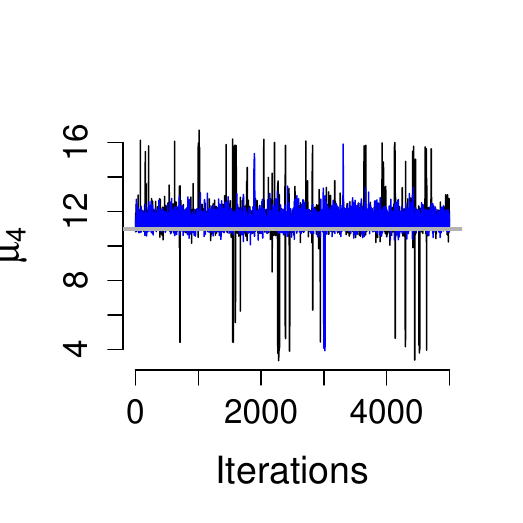}\hspace{-0.3 cm}\includegraphics[scale=0.33]{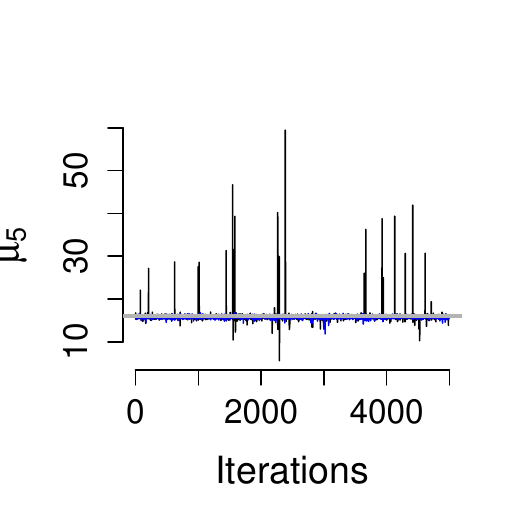}}\\
\hspace{-0.5 cm}\subfigure[Homogeneous - RJMCMC]{\includegraphics[scale=0.33]{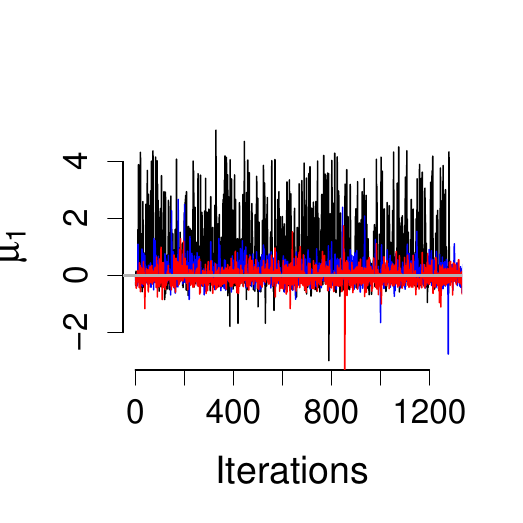}\hspace{-0.3 cm}\includegraphics[scale=0.33]{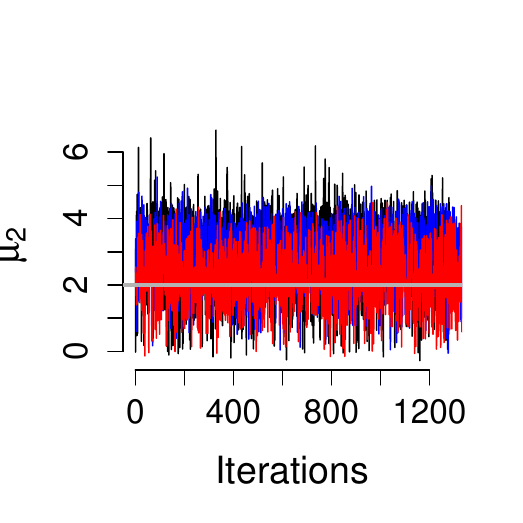}
\hspace{-0.3 cm}\includegraphics[scale=0.33]{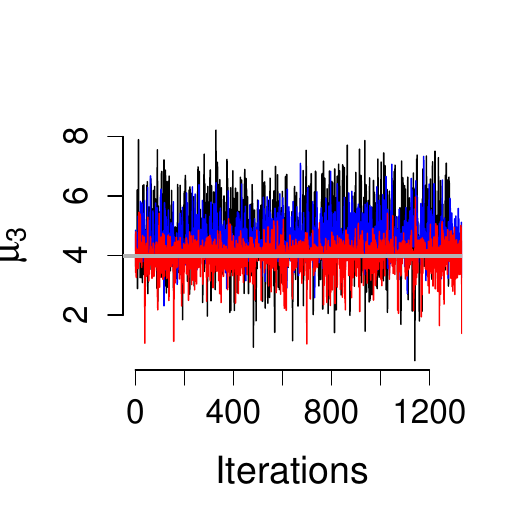}\hspace{-0.3 cm}\includegraphics[scale=0.33]{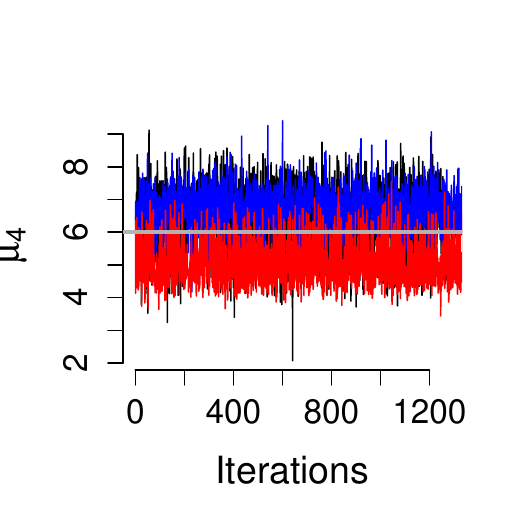}\hspace{-0.3 cm}\includegraphics[scale=0.33]{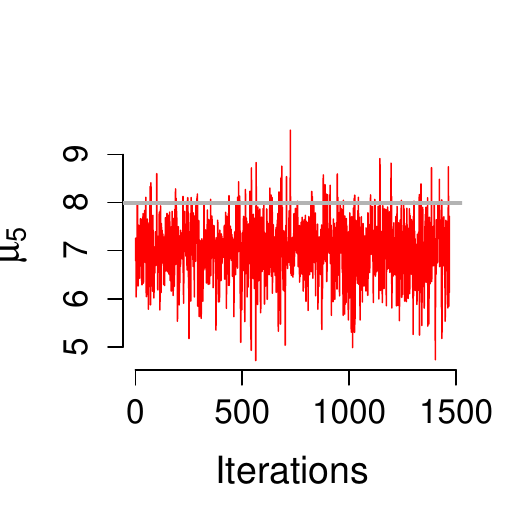}}\\
\hspace{-0.5 cm}\subfigure[Homogeneous - MCMC]{\includegraphics[scale=0.33]{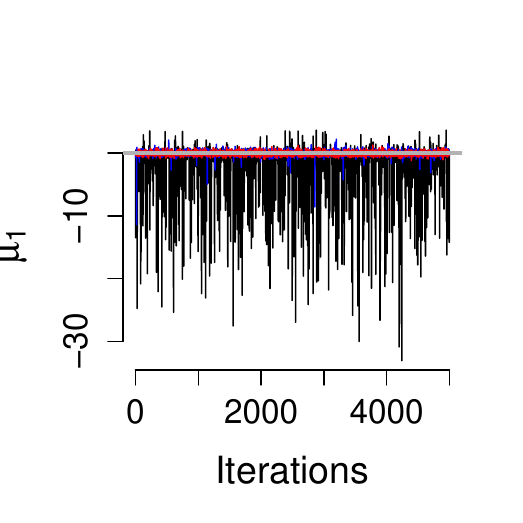}\hspace{-0.3 cm}\includegraphics[scale=0.33]{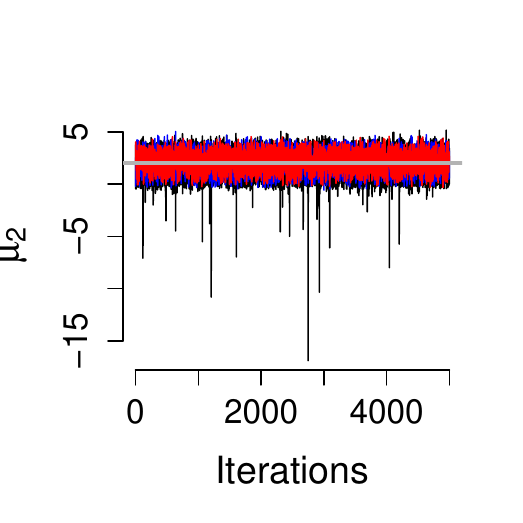}
\hspace{-0.3 cm}\includegraphics[scale=0.33]{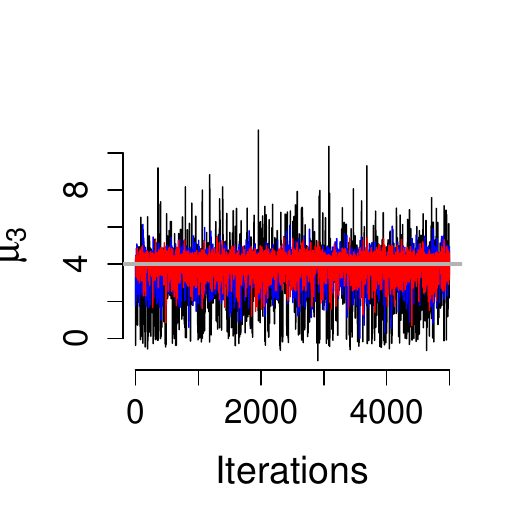}\hspace{-0.3 cm}\includegraphics[scale=0.33]{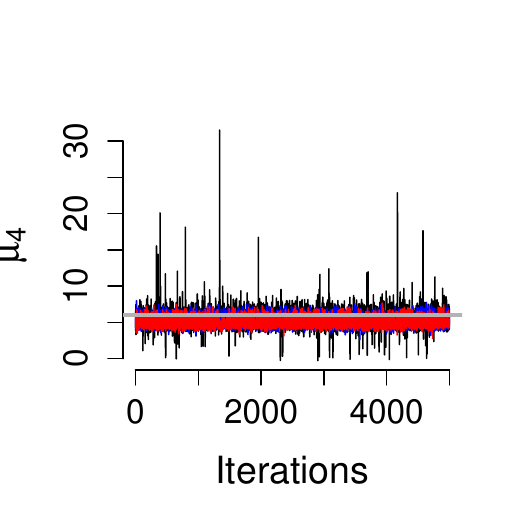}\hspace{-0.3 cm}\includegraphics[scale=0.33]{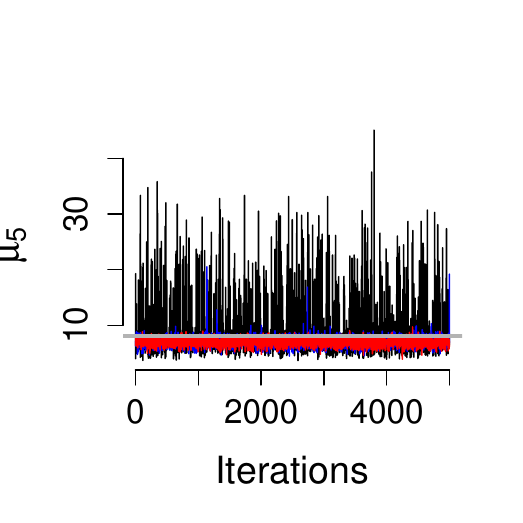}}\\
\end{tabular}
\end{center}
\vspace{-0.3 cm}\caption{{\it Trace plots with the posterior densities of the parameters of $\bfmu$ obtained from fitting the normal 
mixture
model under the different priors considered for ${\bf w}$ and the two samples. We also assumed the value of $k$ to be unknown (RJMCMC) and known (MCMC).
The black trace is obtained assuming $\gamma=1$, the blue one when
$\gamma=4$ and the red with $\gamma=10$. The gray line represents the true value of each $\mu_j$, $j=1,\dots,5$.}}\label{cadeia1}
\end{figure}

Figure \ref{cadeia1} shows the trace plot with the posterior distribution of parameters $\mu_j$ conditional on the posterior samples, whose 
estimated value of $k$ is the one with highest posterior probability. Here we also considered
the value of $k$ known and fixed it as the
true value used to generate the samples, so MCMC was also used to generate samples from the posterior distribution. 
For the MCMC simulations, we generated 70,000 samples from the posterior distribution, discarded the 
first 20,000, and then thinned the chain by taking every 10th sample value, for both datasets generated. 

All the results were obtained for each scenario and value of $\gamma$ 
considered. The black trace represents the posterior density when setting $\gamma=1$, the blue trace when $\gamma=4$ in both scenarios and 
$\gamma=10$ is represented by the red trace in the homogeneous case. The gray line represents the true value of each $\mu_j$.
Note that in the homogeneous case, when RJMCMC is used, there is only a red trace for $\mu_5$. The reason is that the posterior for $k$ favors 
the value 5 only when 
$\gamma=10$. When analyzing Figure \ref{cadeia1}, we can see that the raw samples jump around between different symmetric regions in the 
trace plots, which explains the multimodalities of the marginal densities for raw samples. Thus, the effects of label-switching can be seen in 
the sampled values of the component means for many cases, even in the heterogeneous case. However, this behavior improves when giving some 
prior information about the mixture proportions. It is also possible to observe that in the homogeneous case it is necessary to increase 
the value of $\gamma$ even more, that is, to give more prior information that the proportions observed in groups are similar, in order to 
minimize the label-switching effects and attain the convergence.


The results obtained indicate that the RJMCMC and MCMC chains converged in some cases, but in others the label-switching phenomenon appears 
significantly, so estimating the means on the basis of the RJMCMC and MCMC output is not straightforward. However, as the value of $\gamma$ increases 
this behavior 
improves. If the number of iterations increases, and so does the lag of the chain, the convergence can also improve, but this would require high 
computational effort. Thus, we
suggest careful elicitation of the prior distribution to have better estimates. Almost all the mean parameters are well estimated
when $\gamma=4$ and $\gamma=10$ in the heterogeneous and homogeneous case, respectively. The traces and density estimates for the mixture proportions and variances present this same behavior. 

In general, MCMC and RJMCMC present similar behavior, mainly for the heterogeneous sample generated. A more interesting comparison between
the two approaches is presented in the next subsection.

\begin{figure}[h!]
\begin{center}
\vspace{-0.5 cm}\begin{tabular}{cc}
\hspace{-0.5 cm}\subfigure[Heterogeneous]{\includegraphics[scale=0.65]{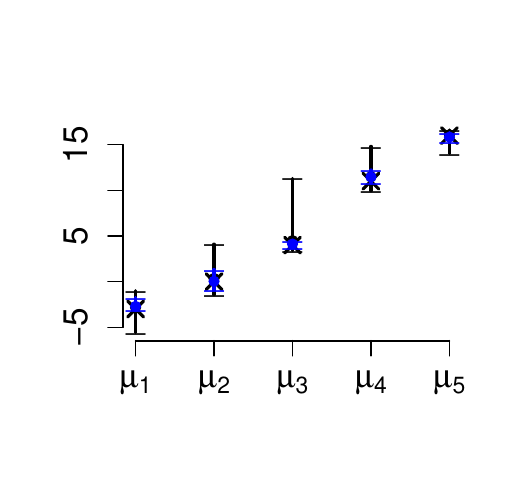}}
\subfigure[Homogeneous]{\includegraphics[scale=0.65]{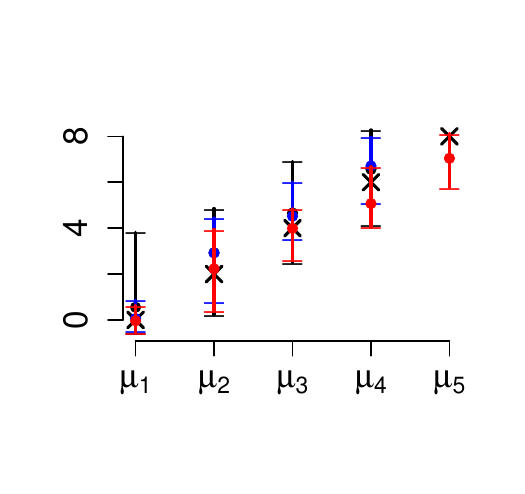}}
\end{tabular}
\end{center}
\vspace{-0.3 cm}\caption{{\it Summary measurements for the point and 95\% credibility interval estimates of the mean parameters model for a 
heterogeneous and homogeneous sample under three prior distributions for ${\bf w}$: the results in black are obtained assuming $\gamma=1$, 
the blue one when $\gamma=4$ and the red when $\gamma=10$. Here it was considered the value of $k$ unknown, so RJCMC was used. 
The crosses represent the true value, the lines the 95\% credibility interval and 
the points are the posterior mean.}}\label{summary1}
\end{figure}

Figure \ref{summary1} shows summary statistics of the posterior distributions of the mean parameters after reaching the supposed 
convergence for each of the scenarios and prior assumed, when assuming $k$ unknown. The crosses represent the true value, the lines the 95\% credibility interval and 
the points are the posterior mean. Also, the results in black are obtained assuming $\gamma=1$, the blue one when $\gamma=4$ and the red when
$\gamma=10$. In almost all the cases, the intervals contain the true value. It is possible to observe the impact of the label-switching, which 
hampers estimating the parameters, but also the improvement in the results when assuming a more informative prior to ${\bf w}$.

Therefore, we conclude that in the cases considered here, the identifiability problem can be minimized under more informative priors and it 
is not necessary to use other alternative approaches to deal with the identifiability problem, like those described in \cite{jasra2005markov}.
Although we named that as an informative prior distribution for the mixture weights, its elicitation does not necessarily need prior information about the weights. In practice, we elicited its hyperparameters based on the MCMC performance and the distribution of the 
dataset among the groups.

The prior distribution of ${\bf w}$ seems to have strong impact on the posterior distribution, improving the results, even in the homogeneous 
case. Furthermore, as the degree of heterogeneity increases, the mixture model's performance improves considerably even under less prior 
information. The same conclusions were attained when estimating the value of $k$ or considering it known.

\begin{figure}[h!]
\begin{center}
\vspace{-0.7 cm}\begin{tabular}{cc}
\hspace{-0.5 cm}\subfigure[Heterogeneous]{\includegraphics[scale=0.6]{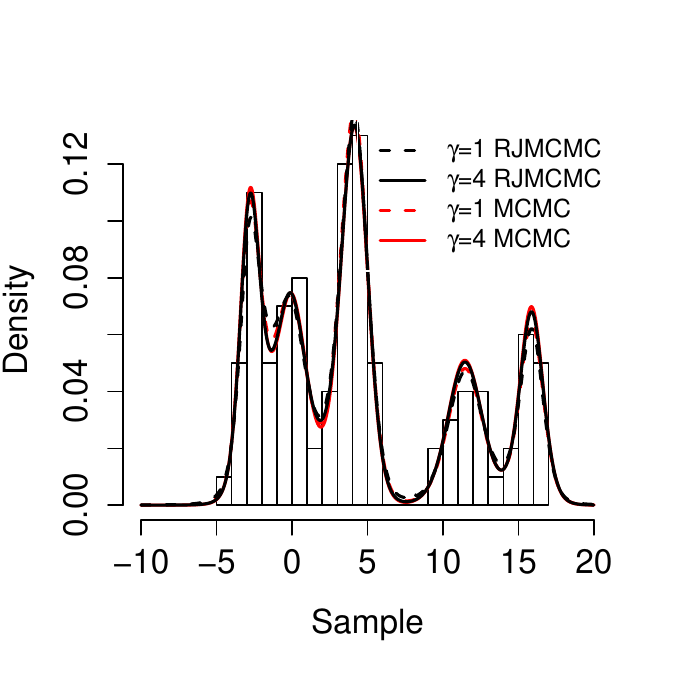}}
\subfigure[Homogeneous]{\includegraphics[scale=0.6]{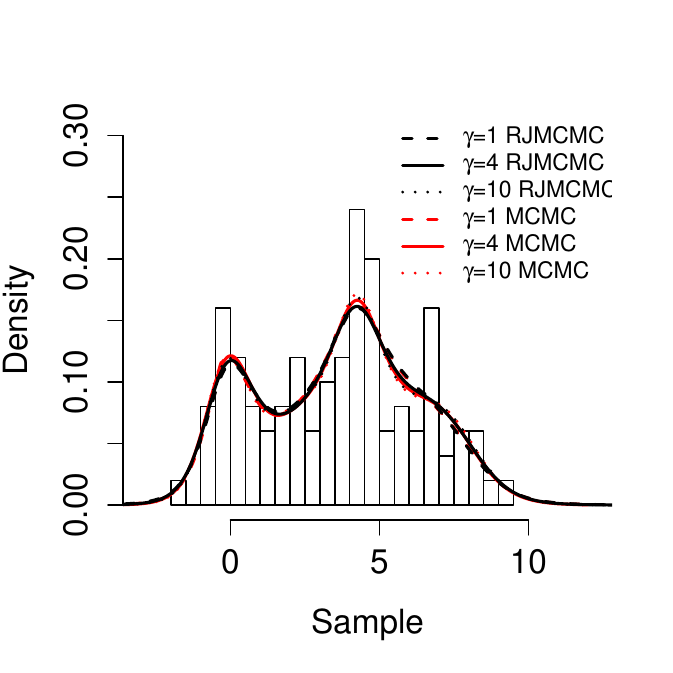}}
\end{tabular}
\end{center}
\vspace{-0.3 cm}\caption{{\it Predictive densities considering different prior distributions for ${\bf w}$ and estimating the value of $k$ 
(RJMCMC) and fixing it on its true value (MCMC).}}\label{pred1}
\end{figure}

Additionally, Figure \ref{pred1} shows the predictive densities for the two datasets generated, for all the
prior distributions considered for ${\bf w}$, represented by the solid ($\gamma=1$), dashed ($\gamma=4$) and dotted ($\gamma=10$) lines, 
respectively. The predictive densities in black are those obtained when the value of $k$ is estimated, so RJMCMC was used, and the red ones
are obtained when the value of $k$ is fixed at the true value, so MCMC was used. The densities obtained under RJMCMC are conditional on the 
posterior samples,
whose sampled value of $k$ is equal to the value with highest posterior probability among all the samples. In contrast with the above results, an estimate of the 
predictive density based on the RJMCMC and MCMC outputs is unaffected by the 
label-switching problem, since it does not depend on how the components are labeled. The predictive density is better 
estimated in the heterogeneous sample than the homogeneous one and the prior distribution does not affect the estimates. Moreover, the 
results obtained in the estimation considering $k$ unknown and fixed are very similar to each other.

\subsection{Comparison between RJMCMC and MCMC}

To examine the performance of the Bayesian estimators obtained under each simulation method, we generated two artificial samples of size $n=100$, 
fixing $k$ at two different values, $k=3$ and $k=5$, in order to evaluate the results when varying the value of $k$. Then, we obtained samples from 
the posterior distribution of the parametric vector, assuming $k$ 
known (MCMC) and estimating it (RJMCMC). In the MCMC simulation we particularly set $k$ for each case at three different values: we assumed it 
to be $2, 3$ and $4$ for the first sample and $4, 5$ and $6$ for the second one. We assumed here the same prior distribution used in Section 
\ref{sens}. Thus, we are interested in evaluating the method's performance when we fix $k$ as its
true value, or a smaller or a greater value than the true one, and when it is estimated. For the RJMCMC and MCMC simulations, we 
generated 70,000 samples from the posterior distribution, discarded the first 10,000, then thinned the chain by taking every 10th sample value, 
and the convergence was achieved.

\begin{figure}[h!]
\centering
\subfigure[$k=3$]{\includegraphics[scale=0.5]{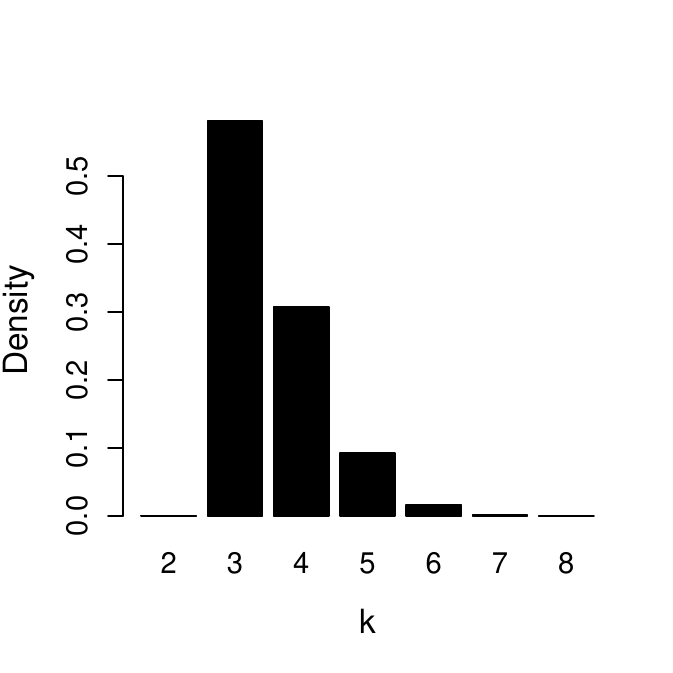}
\includegraphics[scale=0.5]{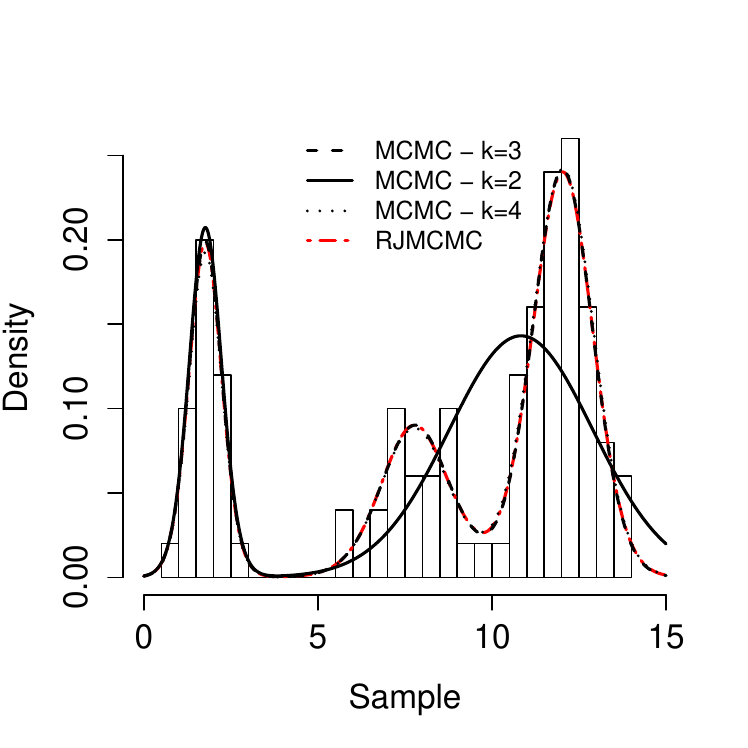}}\\
\subfigure[$k=5$]{\includegraphics[scale=0.5]{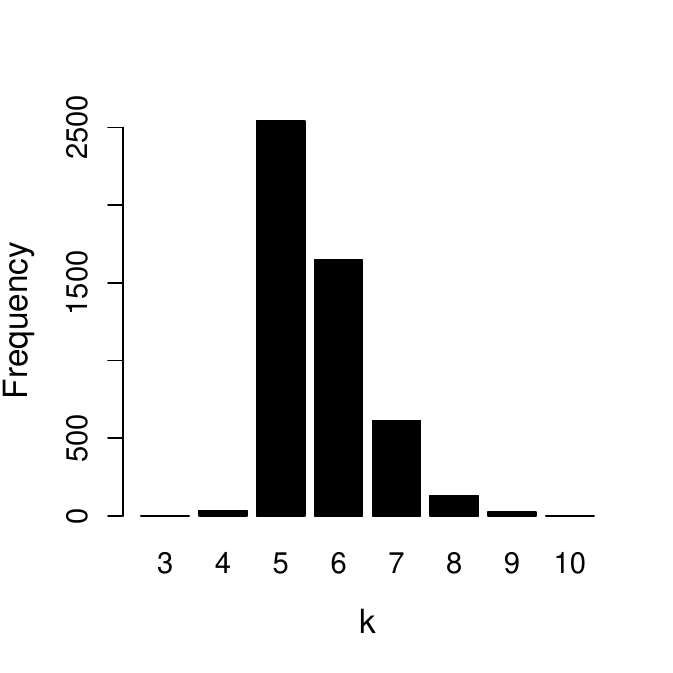}
\includegraphics[scale=0.5]{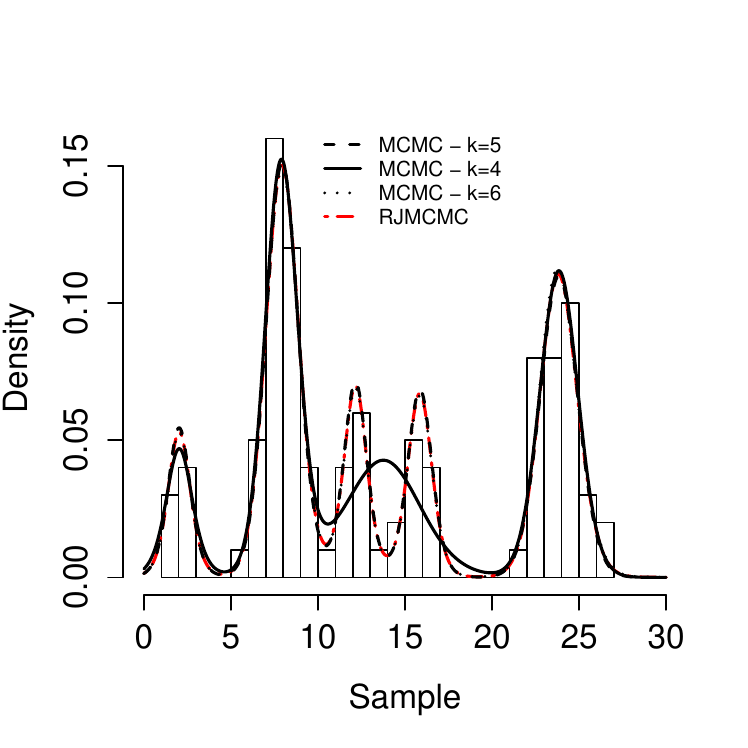}}
\caption{\textit{Predictive densities considering $k$ known (MCMC) and fixing it at $k$, $k-1$, $k+1$ and estimating it (RJMCMC). }}
\label{preditivacomp}
\end{figure}

Figure \ref{preditivacomp} presents the posterior distribution of $k$ obtained in the RJMCMC simulation and the predictive densities 
obtained for each sample generated. It should be noted that $k$ is well estimated and all predictive densities are very similar,
except when we fixed $k$ lower than the true value. Moreover, setting $k$ higher than the true value does not affect the results. 

Selecting a suitable number of mixture components if it is not known in advance, coincides with the problem of model selection. 
A general approach to comparison of complex models based on the samples from the posterior distribution has been suggested by \cite{spiegelhalter2002bayesian} who introduced the Deviance Information Criterion (DIC). Nevertheless, 
its use in mixture models is controversial, because the posterior expectation is not a suitable plug-in estimate for the model parameters 
since it lies in between multiple modes of the posterior density, and alternative plug-in estimators are hard to define.
Other versions of DIC for mixture and in general missing data models have been discussed by \cite{celeux2006deviance}. DIC is computed as
$DIC = \bar{D}+p_D$, such as $p_D=\bar{D}-\tilde{D}$, where $\bar{D}$ is the approximation to the posterior mean of the deviance, 
$\tilde{D}$ is the deviance evaluated in the ``estimate'' to the model parameters and $p_D$ is the effective dimension of parameters.

\cite{plummer2008penalized} suggested to use penalized loss function for Bayesian model comparison and showed
that DIC is an approximation to a penalized loss function based on the deviance, with a penalty derived from a cross-validation argument.
Particularly in mixture context, \cite{plummer2008penalized} recommends to use the Penalized Expected Deviance (PED), which is computed as
$PED = \hat{D}_e+\hat{p}_{opt}$, where $\hat{D}_e$ is the estimated expected deviance, where the estimate is based on two parallel
chains, $\hat{p}_{opt}$ is the estimated penalty, where the estimate is based on simple MCMC average based on two parallel chains.

Table \ref{DIC} and \ref{PED} shows the value of DIC$_3$ presented in \cite{celeux2006deviance} and PED, respectively, for each approach 
considered in this study. As both evaluates the goodness of fit of the model, so the model with the smallest DIC and PED should have the best 
fit. It is possible to observe that, for both criteria, the model with $k$ fixed in the true value seems to fit the data better than its 
counterparts. However, the results are very similar, even when $k$ is 
estimated, increasing the size of the parametric vector, except when $k$ is fixed below the true value. The same conclusion is observed 
for both value fixed for $k$.

\begin{table*}[h!]\caption{\textit{DIC measurements for each model considered.}}\vspace{-0.3 cm}
\hspace{-1 cm}\begin{center}
\begin{tabular}{c|ccc|cccccc} \hline
&\multicolumn{3}{c|}{$k=3$}&\multicolumn{3}{|c}{$k=5$}\\\hline
& DIC & $\bar{D}$ & $p_D$ & DIC & $\bar{D}$ & $p_D$\\\hline 
RJMCMC       & 438.91 & 431.83 & 7.08 & 557.26 & 546.30 & 10.96 \\\hline
MCMC ($k-1$) & 518.09 & 497.03 & 21.06 & 571.38 & 563.39 & 7.99 \\\hline
MCMC ($k$)   & 437.60 & 431.15 & 6.45 & 555.34 & 545.35 & 9.99 \\\hline
MCMC ($k+1$) & 439.80 & 432.29 & 7.51 & 558.12 & 546.53 & 11.59 \\\hline
\end{tabular}\label{DIC}
\end{center}
\end{table*}

\begin{table*}[h!]\caption{\textit{PED measurements for each model considered.}}\vspace{-0.3 cm}
\hspace{-1 cm}\begin{center}
\begin{tabular}{c|ccc|cccccc} \hline
&\multicolumn{3}{c|}{$k=3$}&\multicolumn{3}{|c}{$k=5$}\\\hline
& PED & $\hat{D}_e$ & $\hat{p}_{opt}$ & PED & $\hat{D}_e$ & $\hat{p}_{opt}$ \\\hline 
RJMCMC       & 448.30 & 431.29 & 17.00 & 588.50 & 545.73 & 42.76 \\\hline
MCMC ($k-1$) & 642.81 & 486.45 & 156.36 & 585.34 & 563.43 & 21.91 \\\hline
MCMC ($k$)   & 447.20 & 431.05 & 16.15 & 573.27 & 545.30 & 27.97 \\\hline
MCMC ($k+1$) & 452.14 & 432.26 & 19.88 & 579.22 & 546.11 & 33.11 \\\hline
\end{tabular}\label{PED}
\end{center}
\end{table*}

Table \ref{Comptime} shows the computational time in seconds spent in the fit of each model. As expected, RJMCMC requires 
more computational effort than its counterparts, althought the difference is not so significant. Furthermore, the computational time increases
as $k$ is fixed in a higher value, except when $k$ is estimate.

\begin{table*}[h!]\caption{\textit{Computational time in seconds spent in each models' fit.}}\vspace{-0.3 cm}
\hspace{-1 cm}\begin{center}
\begin{tabular}{c|c|c|c|cccccc} \hline
& RJMCMC & MCMC ($k-1$) & MCMC ($k$) & MCMC ($k+1$) \\\hline
$k=3$ & 3.63 & 1.38 & 1.99 & 2.77\\\hline
$k=5$ & 3.02 & 1.97 & 2.31 & 2.76\\\hline
\end{tabular}\label{Comptime}
\end{center}
\end{table*}

Thus, if the number of components is unknown and we use the MCMC algorithm to sample from the posterior distribution of the parametric vector,
better results are attained by setting it greater than or equal to the true value. On the other hand, estimating the value of $k$ and using the 
RJMCMC method
is a good alternative in this case, having similar performance to the case when we fixed $k$ at its true value.

Finally, we also generated 1,000 samples fixing the parameters at the previous values and obtained samples 
from the posterior distribution of the parametric vector, supposing $k$ known and fixed at the true value in the MCMC and estimating it using
the RJMCMC algorithm. The estimates were then compared with the true values to evaluate the model's performance.

First, in 89.9\% of the 1,000 samples the value of $k$ was correctly estimated when using RJCMC to sample from the posterior distribution. 
Table \ref{tabela1} shows summary statistics with some frequentist measures of the posterior distribution of the model parameters after 
reaching convergence. It reports the square root of the mean square error (SRMSE), the mean absolute error (MAE), the empirical nominal coverage of the $95\%$ 
credibility intervals measured in percentages (Cov.) and the respective widths averaged over the 1,000 simulations (Wid.). 
In particular, the summary statistics of the components parameters are obtained conditioning on $k$ at the value with highest posterior probability.

The parameters are well estimated in both cases and the results are very similar considering each approach, except the parameters 
$\sigma^2_2$ and $\sigma^2_3$, which were slightly better estimated under the RJMCMC and MCMC approaches, respectively. The coverage of the $95\%$ 
credibility intervals is close to the nominal level. These results indicate that similar results can be achieved considering $k$ unknown
and fixing it at the true value. Although the MCMC algorithm has certain advantages with respect to computational cost compared to
the RJMCMC, the number of components is generally unknown and estimating it can be a practical
interest in the problem. Therefore, the RJMCMC is a reasonable alternative to sample from the posterior distribution in this case.

\begin{table*}[h!]\caption{ {\it Summary measurements for the point and $95\%$ credibility interval estimates of the model parameters over 1000 
simulations considering $k$ unknown (RJMCMC) and known (MCMC).}}\vspace{0 cm}
\begin{center}
\footnotesize\begin{tabular}{c|cccccccccc} \hline
& $\mu_1$ & $\mu_2$ & $\mu_3$ & $\sigma_1^2$ & $\sigma_2^2$ & $\sigma_3^2$ & $w_1$ & $w_2$ & $w_3$\\\hline
&\multicolumn{9}{c}{RJMCMC}\\\hline
SRMSE & 0.10 & 0.48 & 0.45 & 0.11 & 0.81 & 0.38 & 0.05 & 0.05 & 0.08\\
MAE  & 0.08 & 0.36 & 0.22 & 0.07 & 0.63 & 0.18 & 0.05 & 0.04 & 0.05\\
Cov. (\%) & 96.7 & 94.1 & 90.0 & 92.0 & 96.2 & 97.6 & 100.0 & 99.9 & 90.6\\
Wid. & 0.45 & 1.96 & 0.84 & 0.39 & 4.08 & 1.23 & 0.16 & 0.22 & 0.25 \\\hline
&\multicolumn{9}{c}{MCMC}\\\hline
SRMSE & 0.10 & 0.48 & 0.14 & 0.11 & 0.96 & 0.18 & 0.05 & 0.06 & 0.03\\
MAE  & 0.08 & 0.36 & 0.11 & 0.07 & 0.70 & 0.15 & 0.05 & 0.04 & 0.03\\
Cov. (\%) & 96.9 & 94.0 & 96.1 & 92.0 & 95.8 & 96.9 & 100.0 & 98.9 & 99.9\\
Wid. & 0.45 & 1.96 & 0.59 & 0.40 & 4.19 & 0.81 & 0.16 & 0.23 & 0.25 \\\hline
\end{tabular}\label{tabela1}
\end{center}
\end{table*}

\section{Application to a real dataset}\label{real_data}

We applied the method to a real dataset that concerns antibody levels of Cytomegalovirus (CMV) in 5,126 individuals, both males and females, 
from 6 years to 49 years old. This dataset was extracted from the 2003 - 2004 National Health and Nutrition Examination Survey (NHANES)\footnote{Centers for Disease Control and Prevention (CDC). National Center for Health Statistics (NCHS). National Health and Nutrition Examination Survey Data. Hyattsville, MD: U.S. Department of Health and Human Services, Centers for Disease Control and Prevention, [2003 - 2004][http://www.cdc.gov/nchs/nhanes].}. 

The CMV is a member of the {\it Herpesviridae} family of viruses and according to \cite{kusne1999prevention}, it is a common virus that occurs 
widely throughout the population but rarely causes noticeable symptoms or significant health problems. 

One method of detecting a CMV infection is by antibody testing of blood samples. This can also be used to determine if someone has had 
recent or past exposure. There are two types of CMV antibodies that are produced in response to a CMV infection, IgM and IgG, and one or both 
can be detected in the blood. IgM antibodies are the first to be produced by the body in response to a CMV infection and they are present in
most individuals within a week or two after the initial exposure. On the other hand, IgG antibodies are produced by the body several weeks 
after the initial CMV infection and provide protection from primary infections. Levels of IgG rise during the active infection, then stabilize 
as the CMV infection resolves and the virus becomes inactive. After a person has been exposed to CMV, the person will have some measurable 
amount of CMV IgG antibodies in his/her blood for the rest of the lifetime. CMV IgG antibody testing can be used, along with IgM testing, 
to help confirm the presence of a recent or previous CMV infection. Particularly, this dataset consists of the IgG levels of CMV. 

\begin{figure}[h!]
\centering
\includegraphics[scale=0.3]{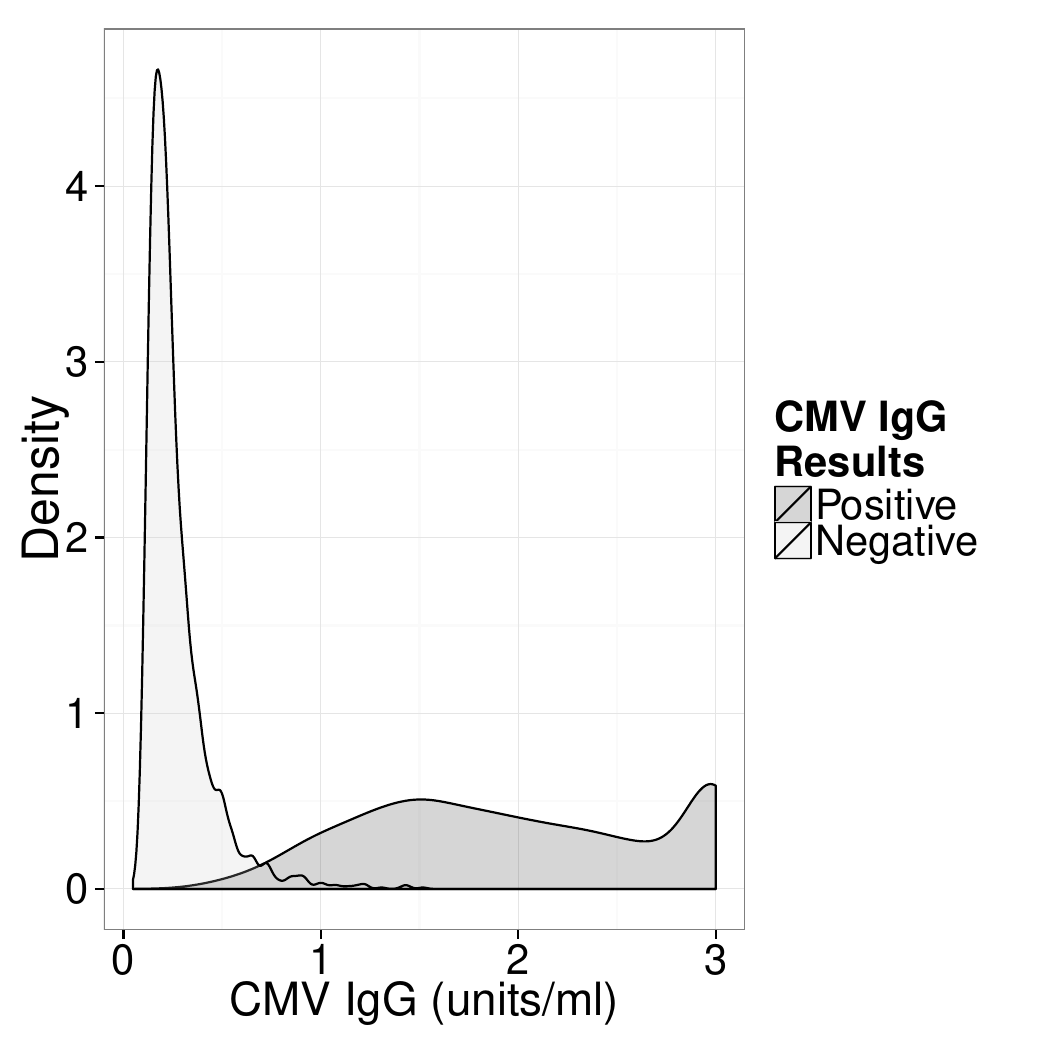}
\caption{\textit{Distribution of antibody levels of CMV IgG (units/ml) for 5,126 individuals infected by the virus or not.}}
\label{real}
\end{figure}

The range of values for the antibody levels CMV 
IgG are from 0.048 to 3.001. For the values reported as ``out of range'' (i.e. over the detectable range, $>3.00$) the survey specialists 
usually assign the value of 3.001. Thus, the observations are left-censored at 3.001 and there are many individuals with this particular value in the dataset. Figure \ref{real} shows the antibody 
levels of CMV IgG distribution for 5,126 individuals infected and not infected. The interest here is in identifying subgroups of IgG as a marker 
of the presence of the disease.

As shown in Figure \ref{real}, some heterogeneous subpopulations can be identified, so it is reasonable to fit the normal mixture model 
(\ref{mixt}) to this dataset. However, in this particular case it is necessary to incorporate left-censoring. It is done
assuming that we observe also $(l_i,u_i)$ associated to each $y_i$, for $i=1,\dots,n$. Note that $y_i=l_i=u_i$ if the observation is not 
censored, $-\infty<l_i<u_i<\infty$ indicates interval-censored observation, $-\infty<l_i<u_i=\infty$ indicates right-censored 
observation and $-\infty=l_i<u_i<\infty$ indicates left-censored observation, which is our particular case. It will be treated in the 
inference, through the likelihood function, which will be written now including the censor to the expression (\ref{likel1}) in the following way:
\begin{eqnarray*}
p({\bf y}|\bfPhi) = \prod_{i=1}^{n} \int_{l_i}^{u_i} \sum_{j=1}^{k} w_{j} f(y_{i}|\bftheta_{j}),
\end{eqnarray*}
with the convention that $\int_l^uf(y)dy=f(l)=f(u)$ whenever $l=u$ (uncensored observation).


Regarding inference, we also considered both estimating $k$ and fixing it on three different values, that are 2, 3 and 4. 
We worked with a subsample of size 1000 selected from the complete data.
For the RJMCMC and MCMC simulations, we generated 50,000 samples from the posterior distribution, discarded the first 10,000, then 
thinned the chain by taking every 10th sample value. Figures \ref{cadeiareal1}, \ref{cadeiareal2}, \ref{cadeiareal3} and \ref{cadeiareal4} in
Appendix B presents the trace plot with the posterior distribution of 
the components of $\bfmu$ and $\bfsigma^2$ for each MCMC and RJMCMC simulations. Analyzing them leads us to conclude that convergence 
seems to be reached for all the parameters. 

\begin{figure}[h!]
\begin{center}
\vspace{-0.3 cm}\begin{tabular}{cc}
\includegraphics[scale=0.5]{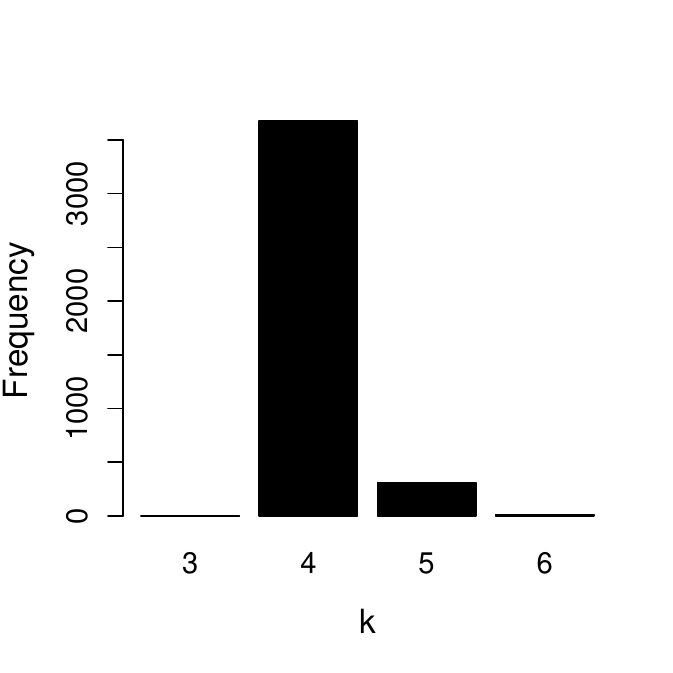}&\includegraphics[scale=0.5]{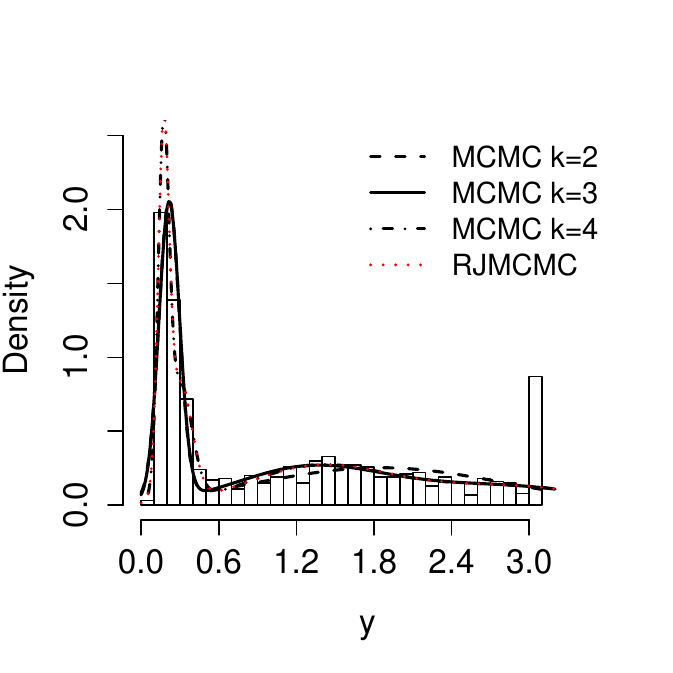}
\end{tabular}
\end{center}
\vspace{-0.5 cm}\caption{{\it Posterior distribution of $k$ and predictive densities for the real dataset.}}\label{pred}
\end{figure}

Figure \ref{pred} displays the posterior distribution of $k$ and predictive densities of antibody levels when estimating $k$ (RJMCMC) and fixing 
it (MCMC), represented by the dashed and dotted lines, respectively. The posterior distribution of $k$ obtained from RJMCMC simulation
favors 4 components. The predictive plots for 
RJMCMC and MCMC with $k$ fixed in 4 are very similar, showing good performance even when $k$ is estimated.    

Table \ref{DIC-PED_real} presents the value of DIC$_3$ and PED for each approach 
considered in this study. Note that, different from previous studies, the model with $k$ unknown presents smaller DIC than the approach
which considered $k$ fixed in 4, although PED be smaller for this one. Thus, using a censored dataset, we observe more benefits in considering
$k$ uknown and using RJMCMC.

\begin{table*}[h!]\caption{\textit{DIC and PED measurements for the models considered to fit the real dataset.}}\vspace{-0.3 cm}
\hspace{-1 cm}\begin{center}
\begin{tabular}{c|ccc|ccc} \hline
& DIC & $\bar{D}$ & $p_D$ & PED & $\hat{D}_e$ & $\hat{p}_{opt}$ \\\hline 
RJMCMC       & 1,826.97 & 1,784.97 & 42.00 & 1,731.00 & 1,696.45 & 34.55   \\\hline
MCMC ($k=2$) & 1,956.46 & 1,912.94 & 43.52 & 1,856.94 & 1,845.74 & 11.19 \\\hline
MCMC ($k=3$)   & 1,916.36 & 1,876.77 & 39.59 & 1,801.89 & 1,774.89 & 27.01 \\\hline
MCMC ($k=4$) & 1,830.53 & 1,788.18 & 42.35 & 1,729.87 & 1,697.75 & 32.12 \\\hline
\end{tabular}\label{DIC-PED_real}
\end{center}
\end{table*}

\section{Conclusions and suggestions for future work}\label{sec:concl}

We considered the problem of the fit of mixture models for heterogeneous populations under 
different levels of heterogeneity. We concluded that in some cases it is not necessary to choose other relabeling algorithms, 
as described in \cite{stephens2000dealing}, in order to 
improve the label-switching problem. To assign a weakly informative prior distribution for the mixture proportions, even for more homogeneous 
populations, was efficient in many cases.

We also evaluated the inference of the model when the number of components
is unknown and RJMCMC maybe used and when it is fixed at a known value. We concluded that when the number of mixture components is unknown,
the RJMCMC is a feasible alternative, achieving similar results when this number is fixed at the true value. Nevertheless, it requires 
slightly greater computational effort than MCMC. On the other hand,
when not interesting in estimating this number, setting it at a value smaller than the true one will generate poor estimates, although, similar
results are obtained when fixing it at the true value or greater than this. We also studied the frequentist properties of the Bayes
estimators obtained from the fits, through a simulation study and we also observed similar results in both approaches.

We applied the methodology to a left-censored real dataset with antibody levels of Cytomegalovirus (CMV) in individuals. 
We concluded here that estimating the value of $k$ was necessary because the distribution of the dataset does not provide much 
information about $4$ subgroups. Thus, it would be possible here to fix $k$ in a small value, and to underestimate the predictive densities. 
Furthermore, different from previous studies, in this case, DIC indicates as the best model the one with $k$ unknown. So, we suggest in 
problems with censored dataset the use of RJMCMC.

Finally, the main findings of this work encourage an extension of this study to other mixture distributions, as the Poisson model discussed in
\cite{viallefont2002bayesian}.

\section*{Acknowledgements}
The authors thank the Associate Editor and the referees for many valuable comments and suggestions.


\begin{thebibliography}{HHH20}
%
%
%
%
%

\bibitem{celeux2006deviance}Celeux, G., Forbes, F., Robert, C. P., Titterington, D. M. et al. (2006) Deviance
information criteria for missing data models. Bayesian analysis, 1, 651-673.

\bibitem{dellaportas2006multivariate}Dellaportas, P. and Papageorgiou, I. (2006) Multivariate mixtures of normals with
unknown number of components. Statistics and Computing, 16, 57-68.
\bibitem{diebolt1994estimation}Diebolt, J. and Robert, C. P. (1994) Estimation of finite mixture
distributions through bayesian sampling. Journal of the Royal
Statistical Society. Series B (Methodological), 363-375.

\bibitem{green1995reversible}Green, P. (1995) Reversible jump markov chain monte carlo computation and bayesian
model determination. Biometrika, {\bf 82}, 711-732.

\bibitem{jasra2005markov}Jasra, A., Holmes, C. and Stephens, D. (2005) Markov chain monte
carlo methods and the label switching problem in bayesian mixture
modeling. Statistical Science, 50-67.

\bibitem{jordan2004graphical}Jordan, M. I. (2004) Graphical models. Statistical Science, 140-155.

\bibitem{komarek2009new}   Komarék, A. (2009) A new r package for bayesian estimation of multivariate normal
mixtures allowing for selection of the number of components and interval-censored data. Computational Statistics \& Data Analysis, 53, 3932-3947.

\bibitem{kusne1999prevention}Kusne, S., Shapiro, R. and Fung, J. (1999) Prevention and treatment
of cytomegalovirus infection in organ transplant recipients. Transplant
infectious disease, {\bf 1}, 187-203.

\bibitem{mclachlan2004finite}McLachlan, G. and Peel, D. (2004) Finite mixture models. John Wiley \& Sons.
\bibitem{nobile2004posterior}Nobile, A. (2004) On the posterior distribution of the number of components in a finite
mixture. Annals of statistics, 2044-2073.

\bibitem{plummer2008penalized} Plummer, M. (2008) Penalized loss functions for bayesian model comparison. Biostatistics,
9, 523-539.

\bibitem{teamR} R Core Team (2014) R: A Language and Environment for Statistical Computing. R
Foundation for Statistical Computing, Vienna, Austria. URL: http://www.R-project.
org/.

\bibitem{roeder1997practical}Roeder, K. and Wasserman, L. (1997) Practical bayesian density estimation using mixtures
of normals. Journal of the American Statistical Association, 92, 894-902.

\bibitem{redner1984mixture}Redner, R. A. and Walker, H. F. (1984) Mixture densities, maximum
likelihood and the em algorithm. SIAM review, {\bf 26}, 195-239.

\bibitem{richardson1997bayesian}Richardson, S. and Green, P. (1997) On bayesian analysis of mixtures with an unknown
number of components. Journal of the Royal Statistical Society, Series B, {\bf 59}, 731-792.

\bibitem{spiegelhalter2002bayesian}Spiegelhalter, D. J., Best, N. G., Carlin, B. P. and Van Der Linde,
A. (2002) Bayesian measures of model complexity and fit. Journal of
the Royal Statistical Society: Series B (Statistical Methodology), {\bf 64}, 583-639.

\bibitem{stephens2000dealing} Stephens, M. (2000) Dealing with label switching in mixture models.
Journal of the Royal Statistical Society: Series B (Statistical Methodology), {\bf 62}, 795-809.


\bibitem{viallefont2002bayesian}Viallefont, V., Richardson, S. and Green, P. J. (2002) Bayesian analysis of poisson
mixtures. Journal of Nonparametric Statistics, {\bf 14}, 181-202.

\end{thebibliography}


\section*{Appendix A: R code to perform the experiments}

All the previous experiments were done using the R package $\mathtt{mixAK}$. Follow we have the main code needed to reproduce the research paradigm for $\bfmu$ and $\bfsigma^2$. To reproduce it for ${\bf w}$, follow the same steps considered for $\bfsigma^2$.

\subsection*{A.1: R code to reproduce studies in Section \ref{simul_study}}
\begin{verbatim}
# *************************
# Generation of the dataset
# *************************
k <- 5
n <- 100
sigma2 <- c(0.22, 1.95, 0.92, 0.74, 1.13)
w <- c(0.17, 0.21, 0.34, 0.12, 0.16)
mu <- y <- list()
mu[[1]] <- c(-3, 0, 4, 11, 16) # heterogeneous population mean
mu[[2]] <- c(0, 2, 4, 6, 8) # homogeneous population mean
z <- sample(1:k,n,prob=w,replace=TRUE) # weights of the subgroups
for(j in 1:2){y[[j]] <- rnorm(n,mu[[j]][z], sqrt(sigma2[z])) # sample
write.table(y[[j]],paste("y_",j,".txt",sep=""),row.names=F,col.names=F)}

# *******************************
# RJMCMC for heterogeneous sample
# *******************************
library(mixAK); library(plyr)# packages required
y <- read.table("y_1.txt",header=F) # read the dataset previous generated
# Prior distribution
po_med <- (range(y)[1] + range(y)[2])/2 # midpoint of Y
delta <- c(1,4)
kmax <- c(5,10)
prio1 <- list(priorK= "uniform", Kmax=kmax[2], delta=delta[1], 
              priormuQ="independentC", xi=po_med, zeta=2*2, g=0.2)
prio2 <- prio1
prio2$delta <- delta[2]
parRJMCMC <- list(par.u1=c(2, 2),par.u2=c(2, 2),par.u3=c(1, 1))                  

# Number of iterations
nMCMC <- c(burn=2000, keep=5000, thin=10, info=1000) #2,000*10+5,000*10

# Models fitted
mod1 <- NMixMCMC(y0=y, prior=prio1, RJMCMC=parRJMCMC, nMCMC=nMCMC,
                 scale=list(shift=0, scale=1),PED=FALSE)
mod2 <- NMixMCMC(y0=y, prior=prio2, RJMCMC=parRJMCMC, nMCMC=nMCMC,
                 scale=list(shift=0, scale=1),PED=FALSE)

# Posterior samples
postK <- postmu <- postSigma <- postw <- postorder <-list()
postK[[1]]<- mod1$K; postK[[2]] <- mod2$K
postmu[[1]] <- mod1$mu; postmu[[2]] <- mod2$mu
postSigma[[1]] <- mod1$Sigma; postSigma[[2]] <- mod2$Sigma
postw[[1]] <- mod1$w; postw[[2]] <- mod2$w
postorder[[1]] <- mod1$order; postorder[[2]] <- mod2$order # order indeces of
# mixture components defined by a relabeling algorithm  

# Posterior summary
grupomu <- gruposig <- postsigorder <- list()
postmu_condK <- postorder_condK <- list()
for (j in 1:2){
  barplot(prop.table(table(postK[[j]])),xlab="k",col="black")
  mtext(substitute(gamma==delta,list(delta=delta[j])), side=3, adj=1)
  valor_k <- as.numeric(names(table(postK[[j]]))[
    table(postK[[j]])==max(table(postK[[j]]))])
  
  # Take posterior marginals conditionals on the value of k
  acumK <- cumsum(postK[[j]]) 
  acumK_alt <- c(0,acumK[1:length(acumK)])
  acumK_menos1 <- acumK_alt[1:length(acumK_alt)-1]
  
  postmu_condK[[j]] <- postorder_condK[[j]] <- list()   
  for (i in 1:length(acumK_menos1)){liminf <- acumK_alt[i] + 1 
   limsup <- acumK_alt[i+1]  
   postmu_condK[[j]][[i]] <- postmu[[j]][liminf:limsup]
   postorder_condK[[j]][[i]] <- postorder[[j]][liminf:limsup]}
  
  indicemu <- which(sapply(postmu_condK[[j]],length)==valor_k)
  
  grupomu[[j]] <- list()
  for (i in indicemu){
    grupomu[[j]] <- rbind(grupomu[[j]],sort(postmu_condK[[j]][[i]]))}
  # conditional mean for each subgroup
  
  postsigorder[[j]] <- list()
  postsigorder[[j]][[1]] <- postSigma[[j]][1:postK[[j]][1]][
    postorder[[j]][1:postK[[j]][1]]]
  for (i in 2:length(postK[[j]])){postsigorder[[j]][[i]] <- postSigma[[j]][
    (sum(postK[[j]][1:(i-1)])+1):sum(postK[[j]][1:i])][postorder[[j]][(
      sum(postK[[j]][1:(i-1)])+1):sum(postK[[j]][1:i])]]}
  
  indicesig <- which(sapply(postsigorder[[j]],length)==valor_k)
  
  gruposig[[j]] <- list()
  for (i in indicesig){gruposig[[j]] <- rbind(gruposig[[j]],
    postsigorder[[j]][[i]])} # conditional variance for each subgroup
}

layout(matrix(1:(2*valor_k),nrow=2,byrow=F))
for (i in 1:valor_k){
  plot(do.call(rbind,grupomu[[1]][,i]),type="l",ylab=substitute(
    paste(mu[nn]),list(nn=i)))
  lines(do.call(rbind,grupomu[[2]][,i]),type="l",col="blue")
  plot(do.call(rbind,gruposig[[1]][,i]),type="l",ylab=substitute(
    paste(sigma^2[nn]),list(nn=i)))
  lines(do.call(rbind,gruposig[[2]][,i]),type="l",col="blue")}

# Predictive densities
pred1 <- NMixPredDensMarg(mod1,grid=seq(-10,20,l=1000))
pred2 <- NMixPredDensMarg(mod2,grid=seq(-10,20,l=1000))
hist(y[[1]],breaks=20,prob=T,ylab="Density", xlim=c(-10,20),xlab="Sample")
lines(pred1$x$x1,pred1$dens$`1`,lwd=2,lty=2)
lines(pred2$x$x1,pred2$dens$`1`,lwd=2,lty=1)

# DIC analysis
mod1$DIC; mod2$DIC

# PED analysis (run the same model, but switch PED argument from FALSE to TRUE).
mod_PED <- NMixMCMC(y0=y[[1]], prior=prio1, RJMCMC=parRJMCMC, nMCMC=nMCMC,
                    scale=list(shift=0, scale=1),PED=TRUE)
mod_PED$PED

# *****************************
# MCMC for heterogeneous sample
# *****************************

prio1_mcmc <- prio1
prio1_mcmc$Kmax <- kmax[1]
prio1_mcmc$xi <- rep(po_med,kmax[1])
prio1_mcmc$priorK <- "fixed"

prio2_mcmc <- prio1_mcmc
prio2_mcmc$delta <- delta[2]

mod1_mcmc <- NMixMCMC(y0=y, prior=prio1_mcmc, nMCMC=nMCMC,
                      scale=list(shift=0, scale=1),PED=FALSE) 
mod2_mcmc <- NMixMCMC(y0=y, prior=prio2_mcmc, nMCMC=nMCMC,
                      scale=list(shift=0, scale=1),PED=FALSE) 

# Posterior samples
postmu <- postSigma <- postw <- postorder <-list()
postmu[[1]] <- mod1_mcmc$mu; postmu[[2]] <- mod2_mcmc$mu
postSigma[[1]] <- mod1_mcmc$Sigma; postSigma[[2]] <- mod2_mcmc$Sigma
postw[[1]] <- mod1_mcmc$w; postw[[2]] <- mod2_mcmc$w
postorder[[1]] <- mod1_mcmc$order; postorder[[2]] <- mod2_mcmc$order 

grupomu <- gruposig <- list()
for (j in 1:2){
  grupomu[[j]] <- t(apply(postmu[[j]], 1, sort))
  
  sig <- data.frame(postSigma[[j]])
  order <- data.frame(postorder[[j]])
  colnames(sig) <- paste0("X",c(1:kmax[1]))
  colnames(order) <- paste0("X",c(1:kmax[1]))
  
  sig$id <- 1:nrow(sig)
  order$id <- 1:nrow(order)
  sig_order <- rbind(sig, order)
  sig_order <- arrange(sig_order, id)
  
  orders <- function(dataset){
    aux <- unlist(dataset[1, -(kmax[1]+1)])
    order <- unlist(dataset[2, -(kmax[1]+1)])
    aux_order <- data.frame(aux, order)
    aux_orders <- arrange(aux_order, order)
    return(aux_orders$aux)
  }
  gruposig[[j]] <- ddply(sig_order, .(id), orders)[,2:(kmax[1]+1)] 
}

layout(matrix(1:(2*kmax[1]),nrow=2,byrow=F))
for (i in 1:kmax[1]){
  plot(grupomu[[1]][,i],type="l",ylab=substitute(paste(mu[nn]),list(nn=i)))
  lines(grupomu[[2]][,i],type="l",col="blue")
  abline(h=mu[[1]][i],col="gray70")
  plot(gruposig[[1]][,i],type="l",ylab=substitute(paste(sigma^2[nn]),list(nn=i)))
  lines(gruposig[[2]][,i],type="l",col="blue")
  abline(h=sigma2[i],col="gray70")
}

# Predictive densities
pred1_mcmc <- NMixPredDensMarg(mod1_mcmc,grid=seq(-10,20,l=1000))
pred2_mcmc <- NMixPredDensMarg(mod2_mcmc,grid=seq(-10,20,l=1000))

hist(y[[1]],breaks=20,prob=T,ylab="Density", xlim=c(-10,20),xlab="Sample")
lines(pred1_mcmc$x$x1,pred1_mcmc$dens$`1`,lwd=2,lty=2)
lines(pred2_mcmc$x$x1,pred2_mcmc$dens$`1`,lwd=2,lty=1)

mod1_mcmc$DIC; mod2_mcmc$DIC # DIC analysis

mod_PED_mcmc <- NMixMCMC(y0=y[[1]], prior=prio1_mcmc, nMCMC=nMCMC,
                         scale=list(shift=0, scale=1),PED=TRUE) # PED analysis
mod_PED_mcmc$PED
\end{verbatim}
\subsection*{A.2: R code to reproduce studies in Section \ref{real_data}}
\begin{verbatim}
library(foreign)
download.file("http://wwwn.cdc.gov/Nchs/Nhanes/2003-2004/SSCMV_C.XPT", 
"download_SSCMV_C.XPT", mode="wb")
y <- sample(read.xport("download_SSCMV_C.XPT")$SSCMVOD,1000)

censor <- rep(1,length(y))
censor[which(y==3.001)] <- 0

prio <- list(priorK= "uniform", Kmax=10, delta=1, xi=sum(range(y))/2,
              priormuQ="independentC", zeta=2*2, g=0.2)

mod_real <- NMixMCMC(y0=y, y1=rep(3.001,length(y)), 
                    censor=censor, prior=prio, RJMCMC = parRJMCMC, nMCMC=nMCMC,
                    scale=list(shift=0, scale=1), PED=FALSE) #RJMCMC 
\end{verbatim}

\section*{Appendix B: Assessment of MCMC and RJMCMC with real data}\label{conv}
\begin{figure}[h!]
\begin{center}
\vspace{-1.1 cm}\begin{tabular}{l}
\hspace{-0.5 cm}\subfigure[$\mu_1$]{\includegraphics[scale=0.3]{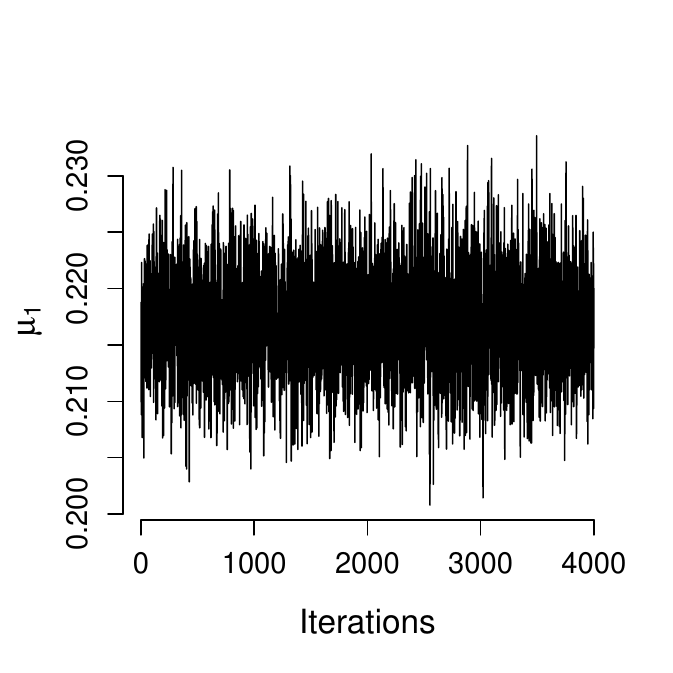}}\hspace{-0.3 cm}\subfigure[$\mu_2$]{\includegraphics[scale=0.3]{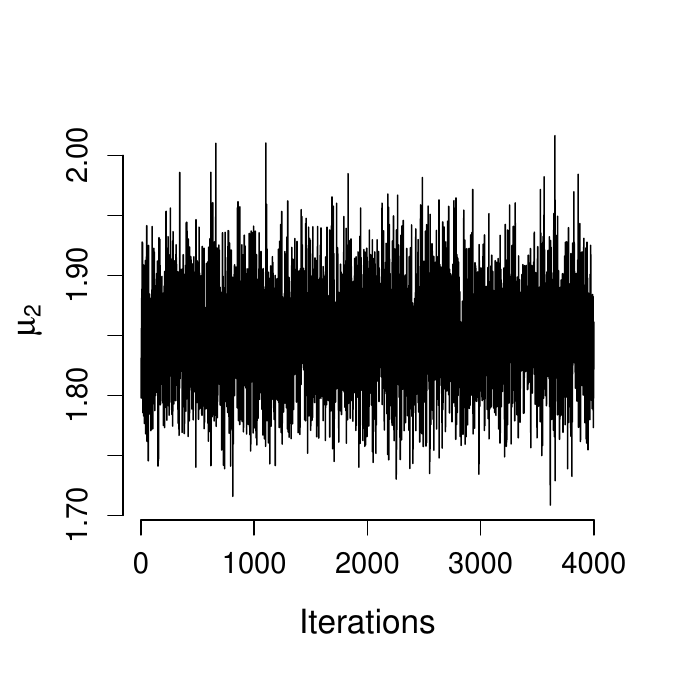}}
\hspace{-0.3 cm}\subfigure[$\sigma_1^2$]{\includegraphics[scale=0.3]{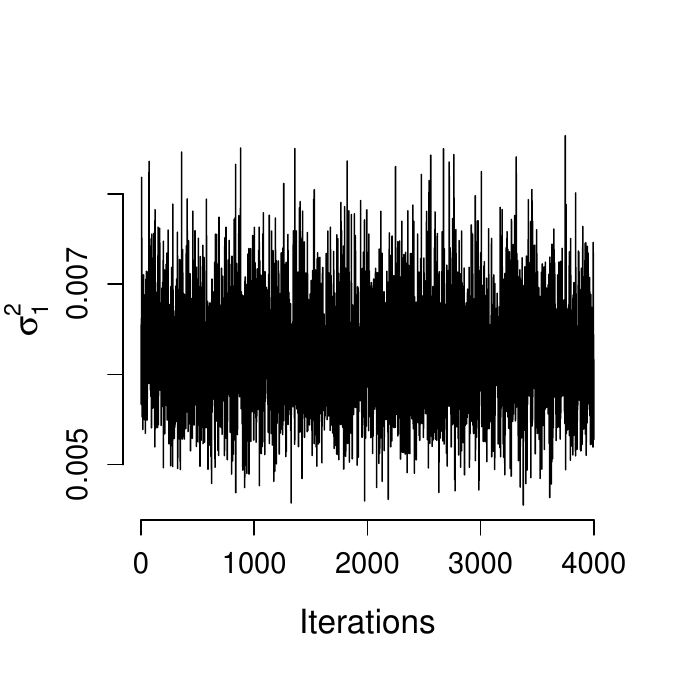}}\hspace{-0.3 cm}\subfigure[$\sigma_2^2$]{\includegraphics[scale=0.3]{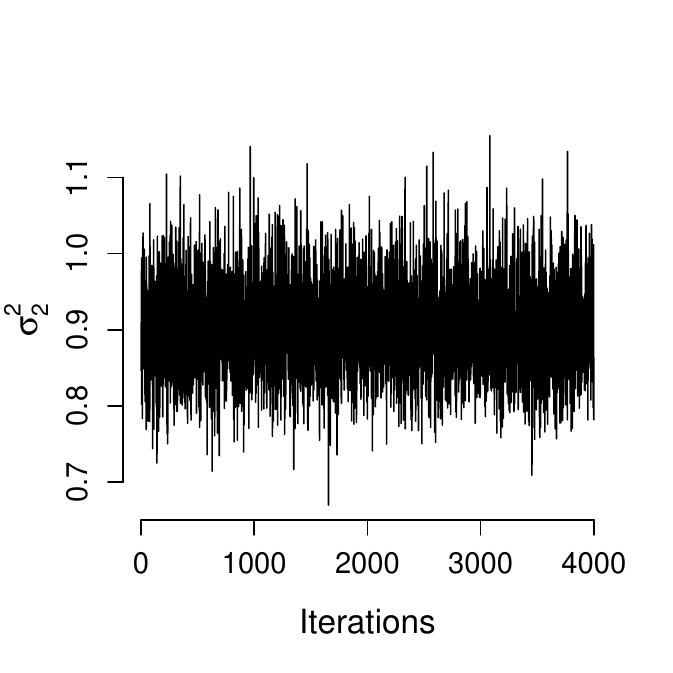}}\\
\end{tabular}
\end{center}
\vspace{-0.3 cm}\caption{{\it Trace plots with the posterior densities of the parameters $\bfmu$ and $\bfsigma^2$ obtained from the fit of the normal mixture model assuming $k$ to be known and fixed at the value 2 in the real dataset (MCMC).}}\label{cadeiareal1}
\end{figure}

\begin{figure}[h!]
\begin{center}
\vspace{-0.7 cm}\begin{tabular}{l}
\hspace{-0.5 cm}\subfigure[$\mu_1$]{\includegraphics[scale=0.3]{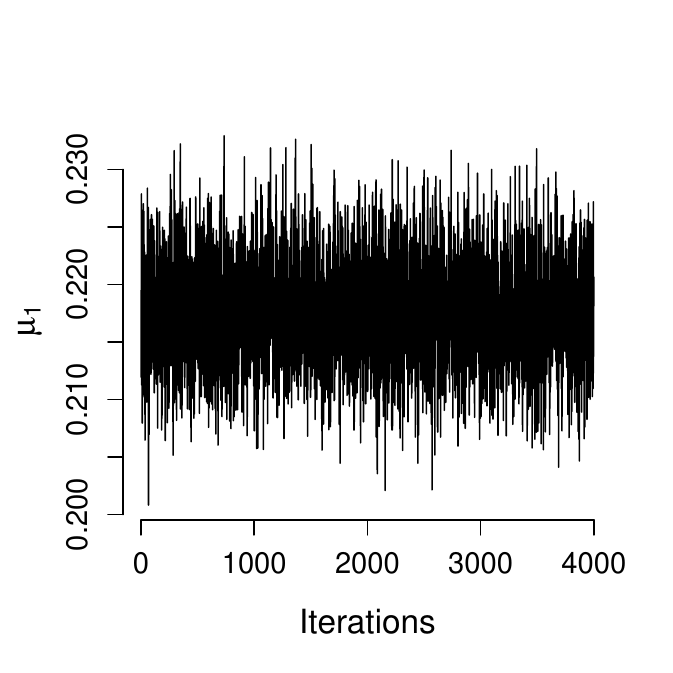}}\hspace{-0.3 cm}\subfigure[$\mu_2$]{\includegraphics[scale=0.3]{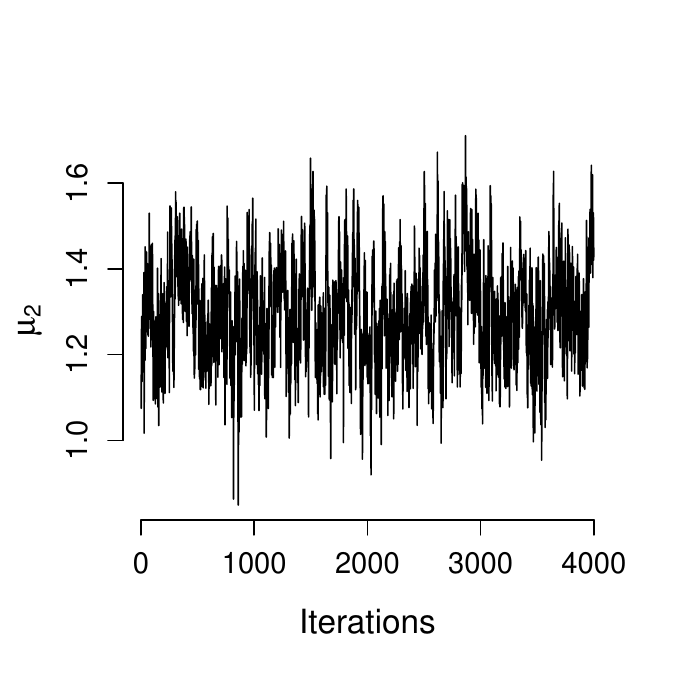}}\hspace{-0.3 cm}\subfigure[$\mu_3$]{\includegraphics[scale=0.3]{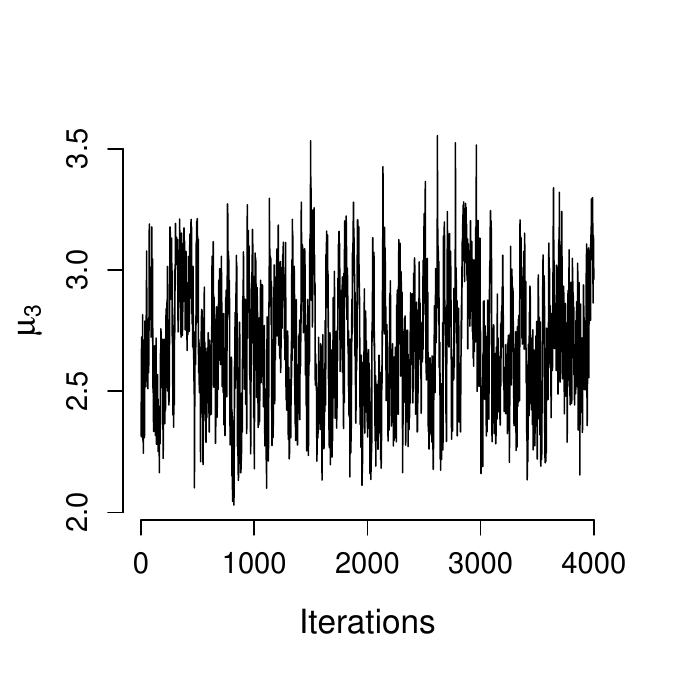}}\\
\hspace{-0.5 cm}\subfigure[$\sigma^2_1$]{\includegraphics[scale=0.3]{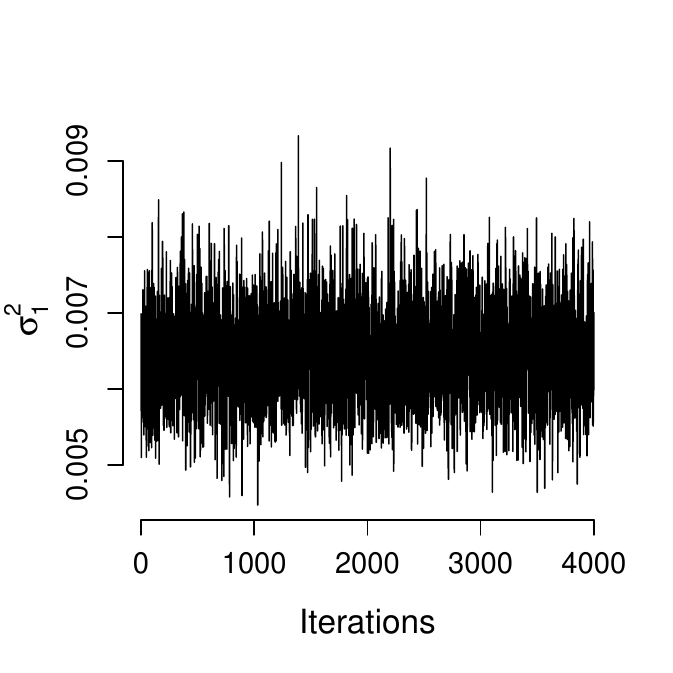}}\hspace{-0.3 cm}\subfigure[$\sigma^2_2$]{\includegraphics[scale=0.3]{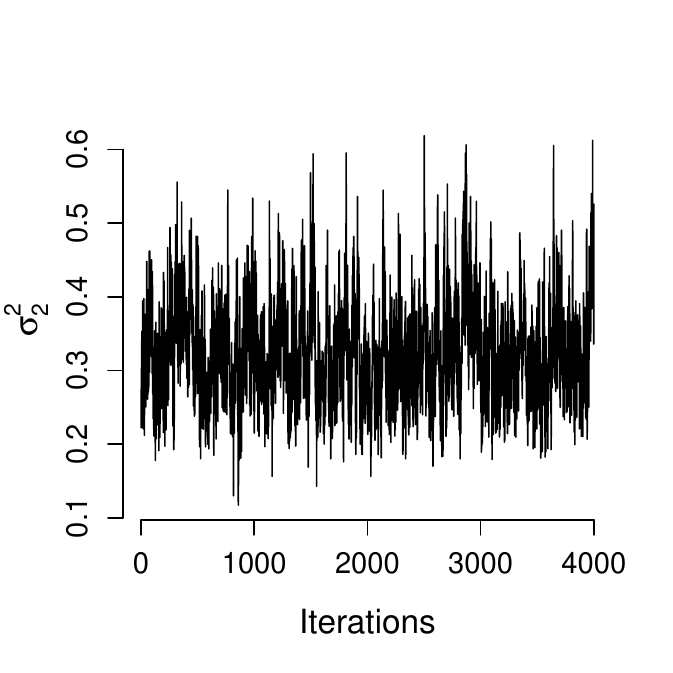}}\hspace{-0.3 cm}\subfigure[$\sigma^2_3$]{\includegraphics[scale=0.3]{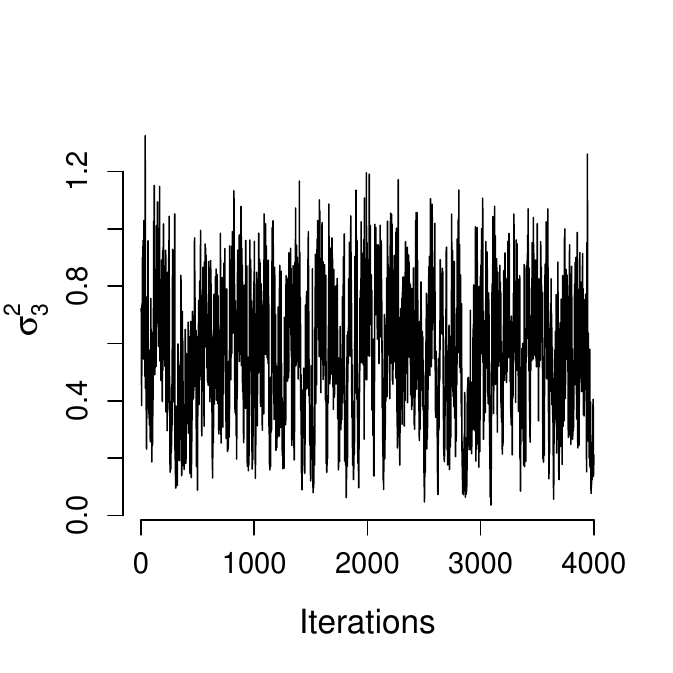}}\\
\end{tabular}
\end{center}
\vspace{-0.3 cm}\caption{{\it Trace plots with the posterior densities of the parameters $\bfmu$ and $\bfsigma^2$ obtained from the fit of the normal mixture model assuming $k$ to be known and fixed at the value 3 in the real dataset (MCMC).}}\label{cadeiareal2}
\end{figure}
\pagebreak

\begin{figure}[h!]
\begin{center}
\vspace{-0.7 cm}\begin{tabular}{l}
\hspace{-0.5 cm}\subfigure[$\mu_1$]{\includegraphics[scale=0.3]{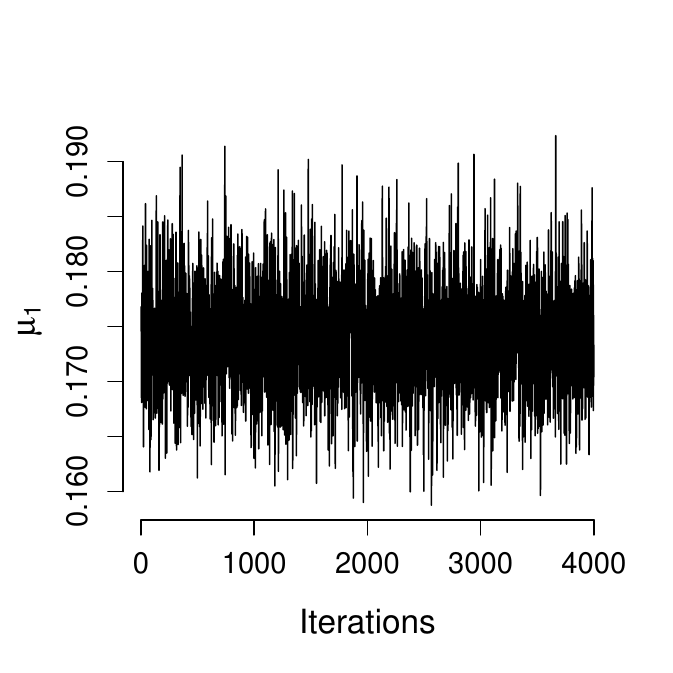}}\hspace{-0.3 cm}\subfigure[$\mu_2$]{\includegraphics[scale=0.3]{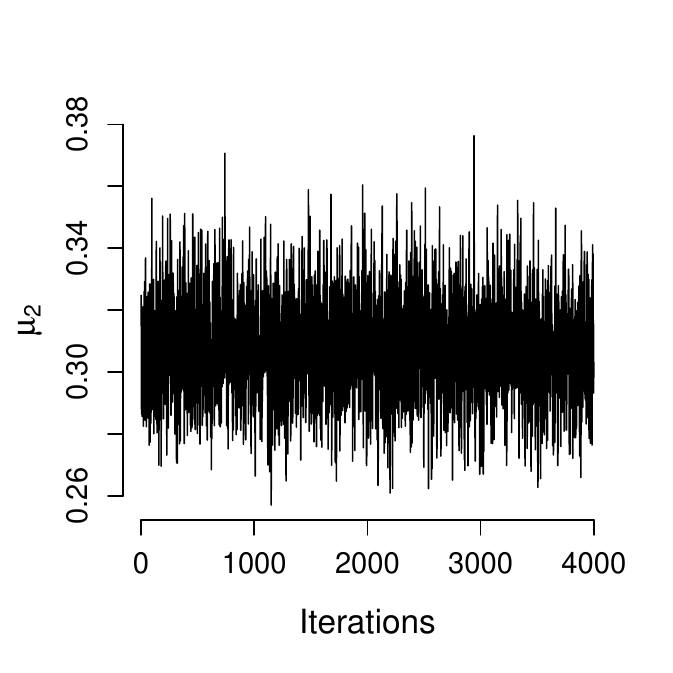}}\hspace{-0.3 cm}\subfigure[$\mu_3$]{\includegraphics[scale=0.3]{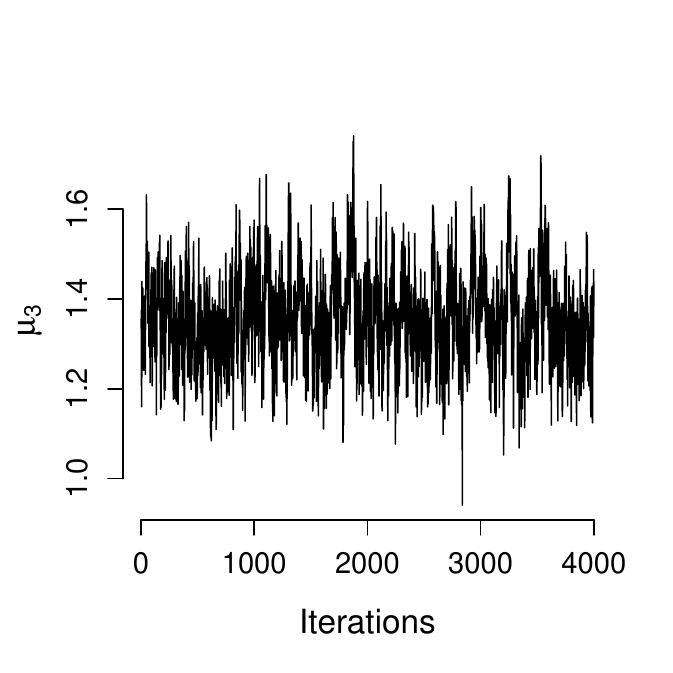}}\hspace{-0.3 cm}\subfigure[$\mu_4$]{\includegraphics[scale=0.3]{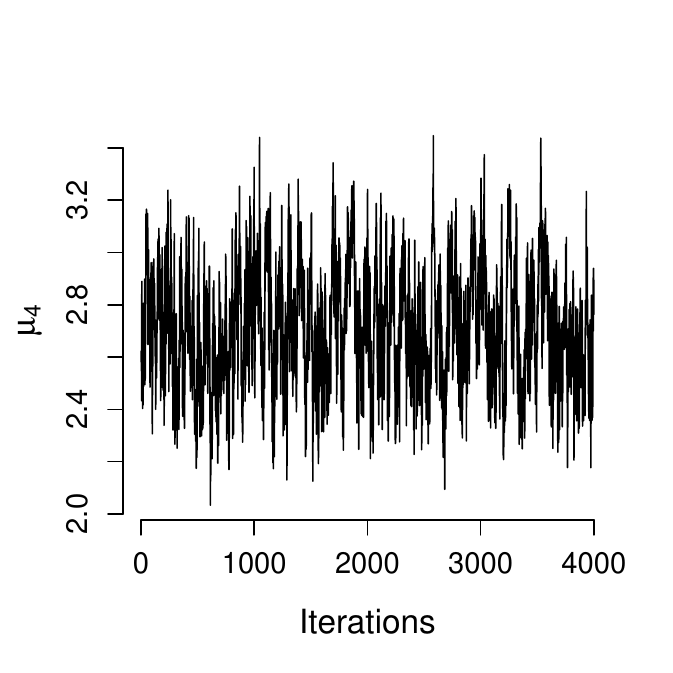}}\\
\hspace{-0.5 cm}\subfigure[$\sigma^2_1$]{\includegraphics[scale=0.3]{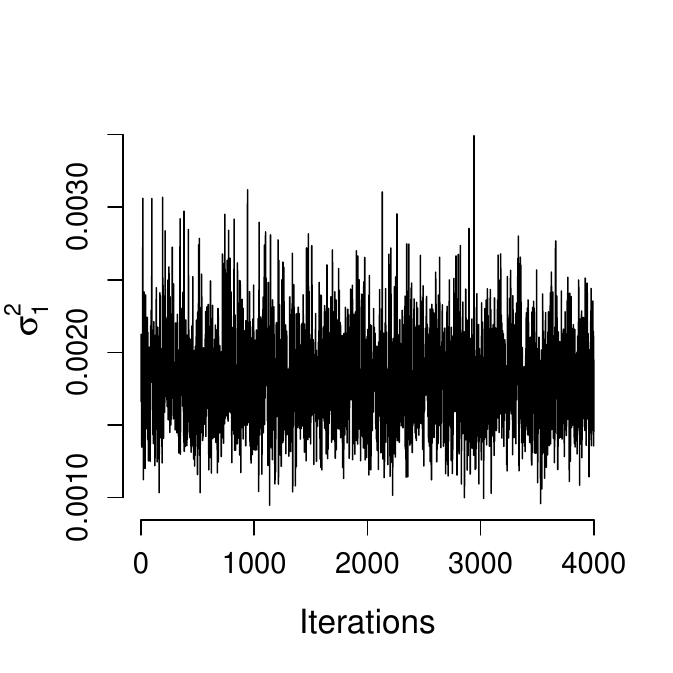}}\hspace{-0.3 cm}\subfigure[$\sigma^2_2$]{\includegraphics[scale=0.3]{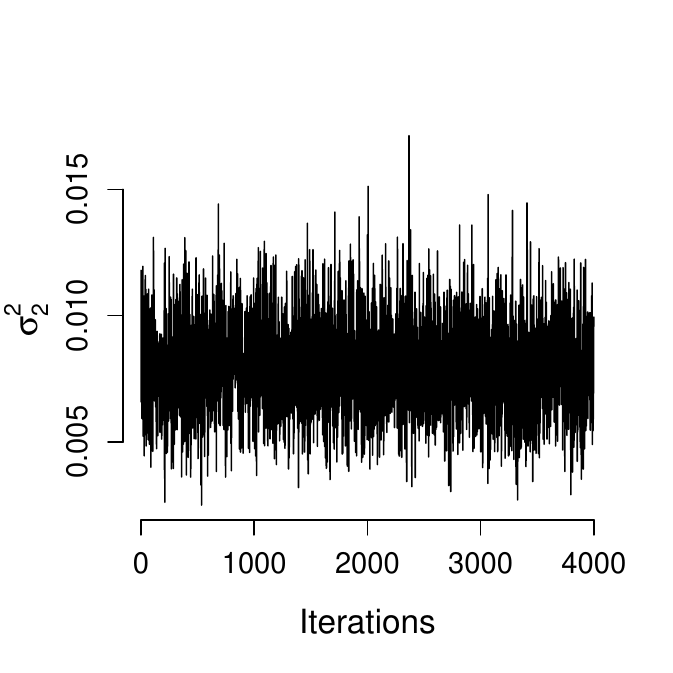}}\hspace{-0.3 cm}\subfigure[$\sigma^2_3$]{\includegraphics[scale=0.3]{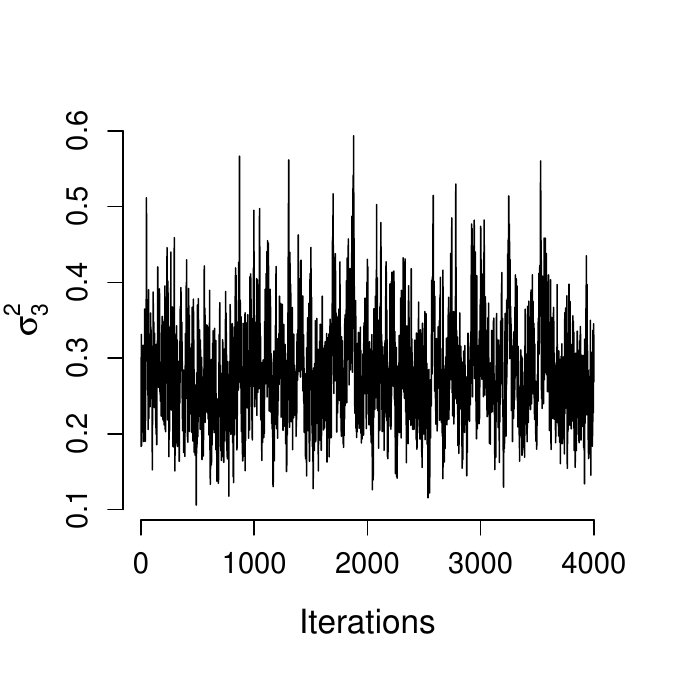}}\hspace{-0.3 cm}\subfigure[$\sigma^2_4$]{\includegraphics[scale=0.3]{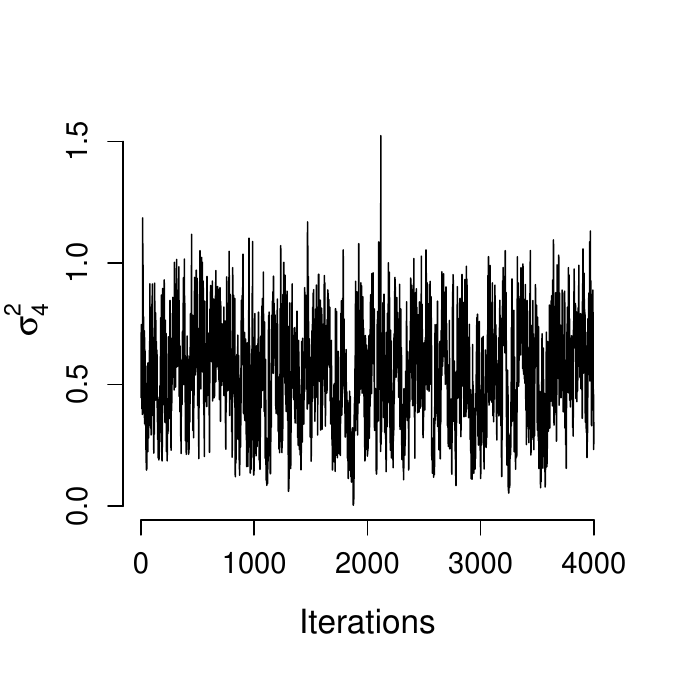}}\\
\end{tabular}
\end{center}
\vspace{-0.3 cm}\caption{{\it Trace plots with the posterior densities of the parameters $\bfmu$ and $\bfsigma^2$ obtained from the fit of the normal mixture model assuming $k$ to be known and fixed at the value 4 in the real dataset (MCMC).}}\label{cadeiareal3}
\end{figure}
\begin{figure}[h!]
\begin{center}
\vspace{-0.1 cm}\begin{tabular}{l}
\hspace{-0.5 cm}\subfigure[$\mu_1$]{\includegraphics[scale=0.3]{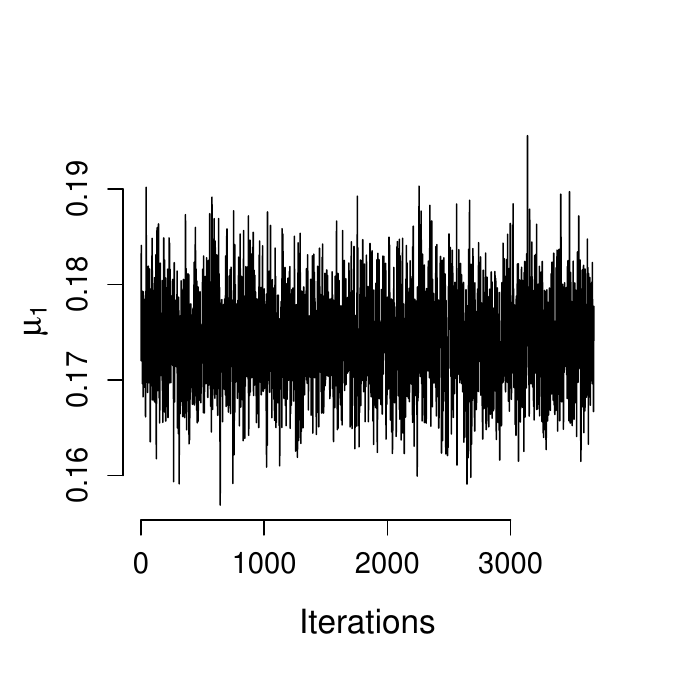}}\hspace{-0.3 cm}\subfigure[$\mu_2$]{\includegraphics[scale=0.3]{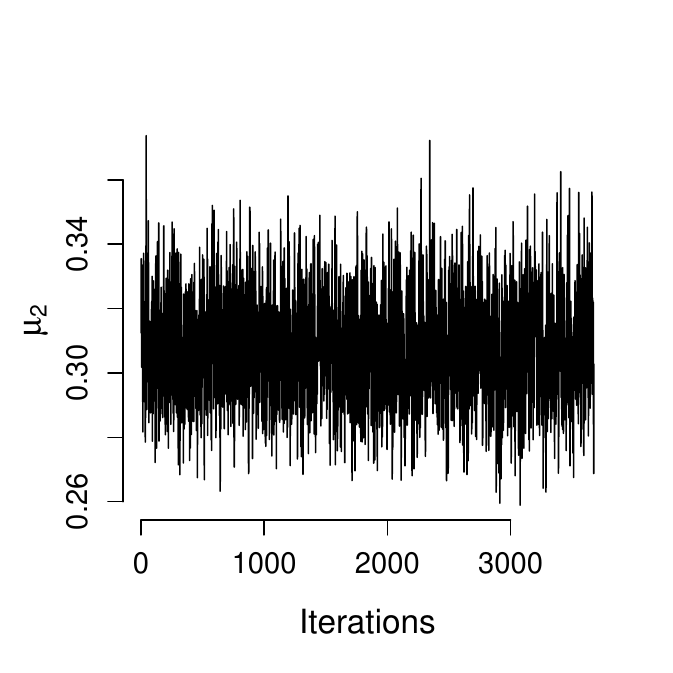}}\hspace{-0.3 cm}\subfigure[$\mu_3$]{\includegraphics[scale=0.3]{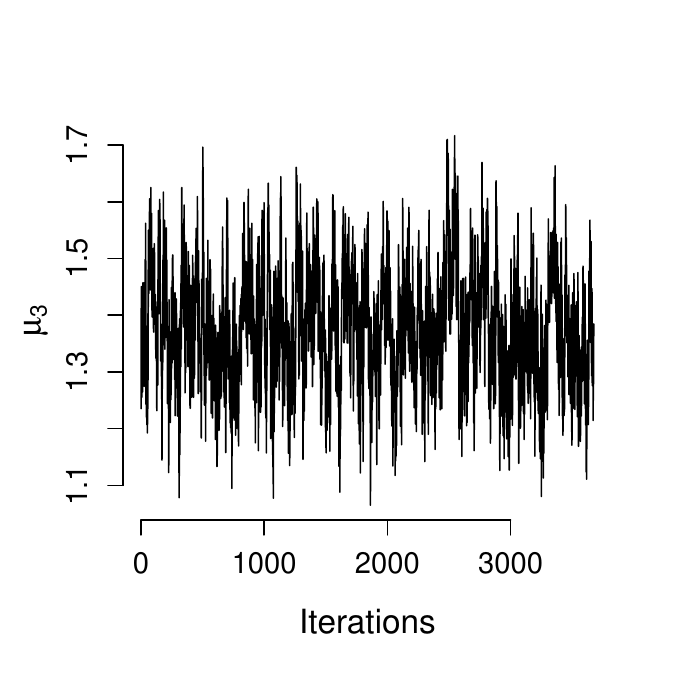}}\hspace{-0.3 cm}\subfigure[$\mu_4$]{\includegraphics[scale=0.3]{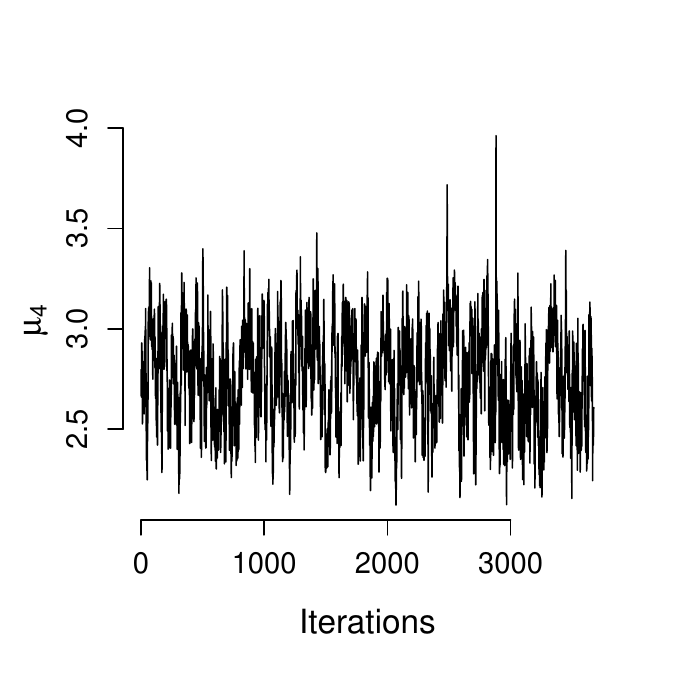}}\\
\hspace{-0.5 cm}\subfigure[$\sigma^2_1$]{\includegraphics[scale=0.3]{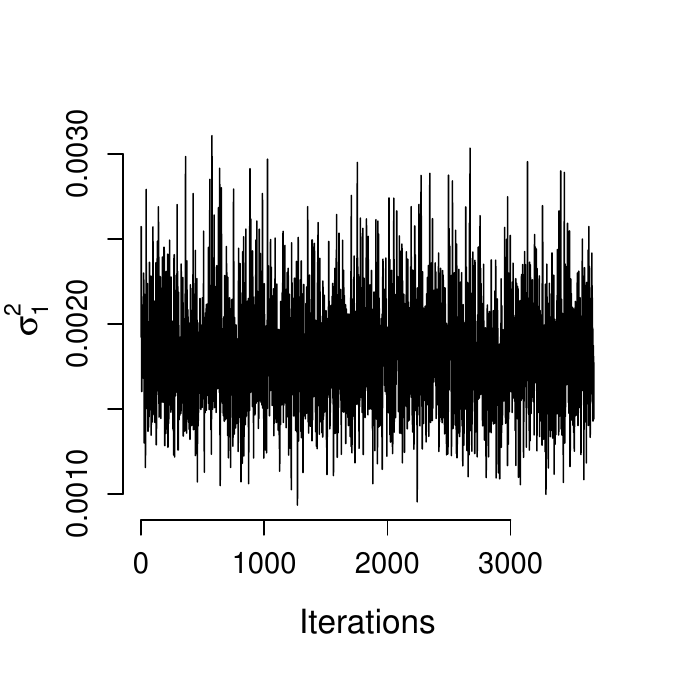}}\hspace{-0.3 cm}\subfigure[$\sigma^2_2$]{\includegraphics[scale=0.3]{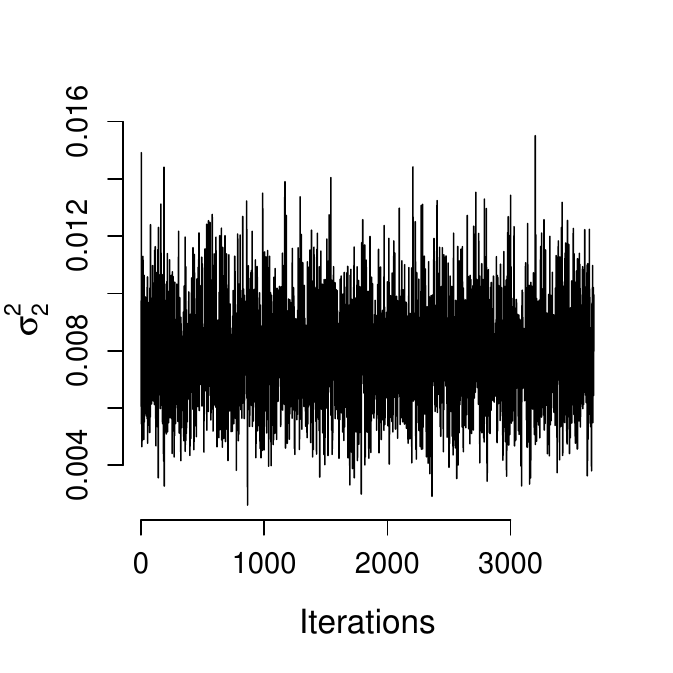}}\hspace{-0.3 cm}\subfigure[$\sigma^2_3$]{\includegraphics[scale=0.3]{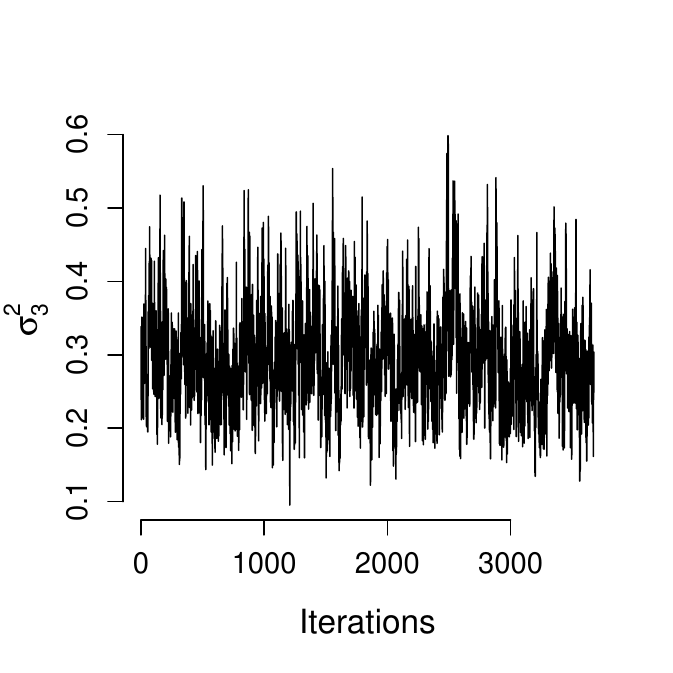}}\hspace{-0.3 cm}\subfigure[$\sigma^2_4$]{\includegraphics[scale=0.3]{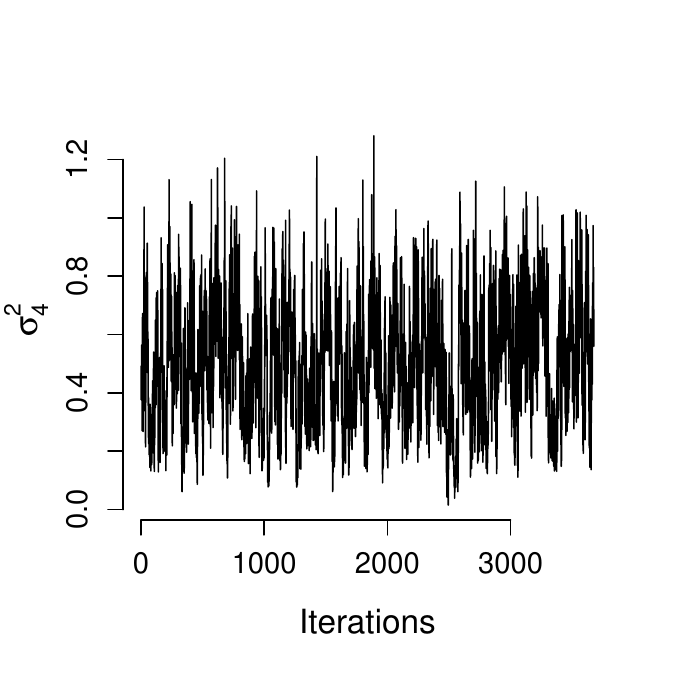}}\\
\end{tabular}
\end{center}
\vspace{-0.3 cm}\caption{{\it Trace plots with the posterior densities of the parameters $\bfmu$ and $\bfsigma^2$ obtained from the fit of 
the normal mixture model assuming $k$ to be unknown in the real dataset (RJMCMC).}}\label{cadeiareal4}
\end{figure}

\end{document}